\documentclass[10pt]{article}
\usepackage{jheppub}
\usepackage{bbold} 
\usepackage[pdftex]{pict2e} 
\allowdisplaybreaks[1]

\newcommand{\ba}{\begin{eqnarray}}
\newcommand{\ea}{\end{eqnarray}}
\newcommand{\la}[1]{\label{#1}}
\newcommand{\fig}{figure~}
\newcommand{\eq}{eq.~}
\newcommand{\se}{section~}
\newcommand{\secs}{sections~}
\newcommand{\app}{appendix~}
\newcommand{\eqs}{eqs.~}
\newcommand{\nr}[1]{(\ref{#1})}
\newcommand{\kallen}{K\"all\'en~}
\newcommand{\intd}[1]{\int\!\!\frac{{\rm d}^d{#1}}{(2\pi)^d}}
\newcommand{\ep}{\varepsilon}

\newcommand{\setN}{\mathbb{N}}
\newcommand{\setZ}{\mathbb{Z}}
\newcommand{\intB}{B}
\newcommand{\intBi}{{\cal B}}
\newcommand{\intJ}{T}
\newcommand{\intG}{P}
\newcommand{\hB}{\intB_{m,m,2m}}
\newcommand{\tB}{\intB_{0,m,m}}
\newcommand{\bintB}{\bar{\intB}}
\newcommand{\bin}[2]{{#1\choose#2}}
\newcommand{\bzmm}[1]{\beta^{#1}}

\newcommand{\Fba}{{}_2F_1}
\newcommand{\Fcb}{{}_3F_2}
\newcommand{\Fdc}{{}_4F_3}
\newcommand{\subA}{\mathbb{1}^{-}}
\newcommand{\subB}{\mathbb{2}^{-}}
\newcommand{\subC}{\mathbb{3}^{-}}

\newcommand{\subDD}{\mathbb{d}^{-}}
\newcommand{\hr}{r}

\newcommand{\po}[2]{\big(#1\big)_{#2}}
\newcommand{\upA}{\mathbb{1}^{+}}
\newcommand{\upB}{\mathbb{2}^{+}}
\newcommand{\upC}{\mathbb{3}^{+}}
\newcommand{\ceil}[1]{\lceil#1\rceil}
\newcommand{\floor}[1]{\lfloor#1\rfloor}
\newcommand{\mabc}{{m_1,m_2,m_3}}
\newcommand{\nabc}{{\nu_1,\nu_2,\nu_3}}
\newcommand{\Nu}{{\Sigma\nu_i}}
\newcommand{\ka}{\tau}
\newcommand{\Xc}{{u}}
\newcommand{\djk}{t_{jk}} 
\newcommand{\barc}{{\tilde{c}}}
\newcommand{\dc}{{d_0}}
\newcommand{\Nn}{{n_0}}
\newcommand{\ct}{{w}}
\newcommand{\ch}{{\tilde\ct}}
\newcommand{\cc}{{\rho}}
\newcommand{\cb}{{\tilde\cc}}
\newcommand{\qp}{{q}}
\newcommand{\hargs}[3]{\Big(\,\substack{#1\\[0.7ex]#2}\;\Big|\,#3\,\Big)}
\newcommand{\hypA}{\Fdc\hargs{\nu_3,\;\nu_1+e,\;\nu_2+e,\;a+e}{(\nu_1+\nu_2)/2+e,\;(\nu_1+\nu_2+1)/2+e,\;1+e}{1}}
\newcommand{\hypB}{\Fdc\hargs{\nu_1,\;\nu_2,\;a,\;d/2}{(\nu_1+\nu_2)/2,\;(\nu_1+\nu_2+1)/2,\;1-e}{1}}

\newcommand{\hypAA}{\Fdc\hargs{a+e,\;\nu_3,\;x_{12}+e,\;n_{12}+e}{f_{12}+e,\;c_{12}+e,\;1+e}{1}}
\newcommand{\hypBB}{\Fdc\hargs{a,\;n_{12},\;x_{12},\;\nu_3-e}{f_{12},\;c_{12},\;1-e}{1}}

\newcommand{\sunB}[3]{\vcenter{\hbox{\begin{picture}(44,50)(-22,-25)
\linethickness{0.5\unitlength} \put(0,0){\circle{40}}
\linethickness{1.5\unitlength} \put(-20,0){\line(1,0){40}}
\put(-20,0){\circle*{3}}
\put(20,0){\circle*{3}}
\put(0,22){\makebox(0,0)[b]{\footnotesize #1}}
\put(0,2){\makebox(0,0)[b]{\footnotesize #3}}
\put(0,-22){\makebox(0,0)[t]{\footnotesize #2}}
\end{picture}}}}
\newcommand{\tads}[4]{\vcenter{\hbox{\begin{picture}(24,50)(-12,-25)
\linethickness{#1\unitlength} \put(0,10){\circle{20}}
\linethickness{#2\unitlength} \put(0,-10){\circle{20}}
\put(0,0){\circle*{3}}
\put(0,22){\makebox(0,0)[b]{\footnotesize #3}}
\put(0,-22){\makebox(0,0)[t]{\footnotesize #4}}
\end{picture}}}}
\newcommand{\tadA}[2]{\tads{0.5}{0.5}{#1}{#2}}
\newcommand{\tadB}[2]{\tads{0.5}{1.5}{#1}{#2}}

\newcommand{\Ja}{\vcenter{\hbox{\begin{picture}(34,40)(-17,-20)
\put(0,0){\circle{20}}
\put(0,10){\line(0,1){10}}
\put(-10,-20){\line(0,1){20}}
\put(10,-20){\line(0,1){20}}
\put(0,10){\circle*{3}}
\put(-10,0){\circle*{3}} \put(10,0){\circle*{3}}
\put(2,15){\makebox(0,0)[l]{\footnotesize $p_3$}}
\put(-11,-15){\makebox(0,0)[r]{\footnotesize $p_1$}}
\put(12,-15){\makebox(0,0)[l]{\footnotesize $p_2$}}
\end{picture}}}}
\newcommand{\Jb}{\vcenter{\hbox{\begin{picture}(34,40)(-17,-20)
\put(0,0){\circle{20}}
\put(10,0){\line(0,1){20}}
\put(-10,-20){\line(0,1){20}}
\put(10,-20){\line(0,1){20}}
\put(-10,0){\circle*{3}} \put(10,0){\circle*{3}}
\put(-11,-15){\makebox(0,0)[r]{\footnotesize $p_1$}}
\end{picture}}}}
\newcommand{\Jc}{\vcenter{\hbox{\begin{picture}(34,40)(-17,-20)
\put(0,0){\circle{20}}
\put(-10,0){\line(0,1){20}}
\put(-10,-20){\line(0,1){20}}
\put(10,-20){\line(0,1){20}}
\put(-10,0){\circle*{3}} \put(10,0){\circle*{3}}
\put(12,-15){\makebox(0,0)[l]{\footnotesize $p_2$}}
\end{picture}}}}
\newcommand{\Jd}{\vcenter{\hbox{\begin{picture}(34,40)(-17,-20)
\put(0,0){\circle{20}}
\put(0,10){\line(0,1){10}}
\put(-1,-20){\line(0,1){10}}
\put(1,-20){\line(0,1){10}}
\put(0,10){\circle*{3}} \put(0,-10){\circle*{3}}
\put(2,15){\makebox(0,0)[l]{\footnotesize $p_3$}}
\end{picture}}}}

\newcommand{\doubleSumSplit}{%
\begin{picture}(400,80)(-10,-12)
\put(-2,0){\vector(1,0){72}}
\put(0,-2){\vector(0,1){72}}
\put(68,-12){$j$}
\put(-10,67){$k$}
\put(8,-12){$1$}
\put(47,-12){$\nu$}
\put(-10,-2){$0$}
\put(-10,48){$\nu$}
\put(-2,50){\line(1,0){4}}
\multiput(10,0)(0,10){2}{\circle*{4}}
\multiput(20,0)(0,10){3}{\circle*{4}}
\multiput(30,0)(0,10){4}{\circle*{4}}
\multiput(40,0)(0,10){5}{\circle*{4}}
\multiput(50,0)(0,10){6}{\circle*{4}}
\put(80,28){$=$}
\put(108,0){\vector(1,0){122}}
\put(110,-2){\vector(0,1){72}}
\put(225,-12){$j$}
\put(100,67){$k$}
\put(118,-12){$1$}
\put(157,-12){$\nu$}
\put(205,-12){$2\nu$}
\put(100,-2){$0$}
\put(100,48){$\nu$}
\put(108,50){\line(1,0){4}}
\multiput(120,0)(0,10){2}{\circle*{4}}
\multiput(130,0)(0,10){3}{\circle*{4}}
\multiput(140,0)(0,10){4}{\circle*{4}}
\multiput(150,0)(0,10){5}{\circle*{4}}
\multiput(160,0)(0,10){6}{\circle*{4}}
\multiput(170,0)(0,10){5}{\circle*{4}}
\multiput(180,0)(0,10){4}{\circle*{4}}
\multiput(190,0)(0,10){3}{\circle*{4}}
\multiput(200,0)(0,10){2}{\circle*{4}}
\multiput(210,0)(0,10){1}{\circle*{4}}
{\color{blue}
\put(110,0){\vector(1,1){60}}
\put(172,62){\scriptsize$\ell$}
\multiput(112,8)(5,5){10}{\line(1,-1){10}}
\put(110,10){\tiny$0$}
\put(115,15){\tiny$1$}
\put(134,35){\tiny$\nu$}
\put(148,55){\tiny$2\nu\!\!-\!\!1$}
}
\put(240,28){$-$}
\put(268,0){\vector(1,0){122}}
\put(270,-2){\vector(0,1){72}}
\put(385,-12){$j$}
\put(260,67){$k$}
\put(278,-12){$1$}
\put(317,-12){$\nu$}
\put(365,-12){$2\nu$}
\put(260,-2){$0$}
\put(260,48){$\nu$}
\put(320,-2){\line(0,1){4}}
\put(280,-2){\line(0,1){4}}
\put(268,50){\line(1,0){4}}
\multiput(330,0)(0,10){5}{\circle*{4}}
\multiput(340,0)(0,10){4}{\circle*{4}}
\multiput(350,0)(0,10){3}{\circle*{4}}
\multiput(360,0)(0,10){2}{\circle*{4}}
\multiput(370,0)(0,10){1}{\circle*{4}}
{\color{blue}
\put(270,0){\vector(1,1){60}}
\put(332,62){\scriptsize$\ell$}
\multiput(272,8)(5,5){10}{\line(1,-1){10}}
\put(270,10){\tiny$0$}
\put(275,15){\tiny$1$}
\put(294,35){\tiny$\nu$}
\put(308,55){\tiny$2\nu\!\!-\!\!1$}
}
\end{picture}}

\title{Recursion-free solution for two-loop vacuum integrals with ``collinear" masses}

\author{Andrei~I.~Davydychev}
\author{and York~Schr\"oder}

\affiliation{Centro de Ciencias Exactas, Depto.\ de Ciencias B\'asicas, Univ.\ del B\'io-B\'io, Chill\'an, Chile}

\emailAdd{adavyd@ubiobio.cl}
\emailAdd{yschroder@ubiobio.cl}


\abstract{We investigate the structure of a particular class of massive vacuum Feynman integrals at two loops. 
This class enjoys the linear relation $m_1+m_2=m_3$ between its three propagator masses, corresponding to zeros of the associated \kallen function.
Apart from having applications in thermal field theory, the integrals can be mapped onto one-loop three-point functions with collinear external momenta, suggesting the term ``collinear" masses. 
We present a closed-form solution for these integrals, proving that they can always be factorized into products of one-loop cases, 
for all integer-valued propagator powers.}

\begin{document}
\maketitle

%
\section{Introduction}
\la{se:intro}

Loop integrals with internal propagator masses play a prominent role in contemporary precision collider phenomenology. 
They are also interesting objects from a mathematical point of view, motivating the study of classes of special functions,
see \cite{Smirnov:2012gma,Weinzierl:2022eaz,Abreu:2022mfk,Blumlein:2022zkr} for recent reviews of techniques and results.
For example, tremendous efforts have been invested into studying massive two-point functions at two loops, the so-called
sunrise (aka sunset aka banana) class of Feynman integrals \cite{Lairez:2022zkj}, 
giving rise to elliptic structures that go beyond the polylogarithmic results \cite{Bourjaily:2022bwx} known 
to arise from a number of simpler two-loop examples.

One is typically interested in resolving integrals with higher propagator powers 
as well, as those arise e.g.\ from diagrammatic insertions, from taking derivatives when evaluating moments of observables
such as for example when expanding two-point \cite{Davydychev:1992mt,Fleischer:1994ef} 
or three-point functions \cite{Davydychev:1995nq} in external momenta,
or from gauge-field propagators. A standard method to reduce such higher propagator powers to lower values is the
systematic exploitation \cite{Laporta:2000dsw} of integration-by-parts (IBP) identities \cite{Chetyrkin:1981qh,Tkachov:1981wb}, 
which provide for a recursive mapping of large sets of Feynman integrals
to a small linearly independent basis of so-called master integrals. 

Oftentimes, however, the IBP reduction has turned out 
to be a bottleneck (note however new developments, connected to intersection 
theory \cite{Mizera:2019ose,Frellesvig:2020qot,Caron-Huot:2021iev}, or methods exploiting relations that involve 
also changes in the space-time dimension \cite{Tarasov:1996br,Lee:2009dh}).
It is therefore extremely valuable to solve specific classes of Feynman integrals analytically, in non-recursive form. 
Such general all-order solutions are rather scarce, exceptions being some massless cases such as two-point functions
with massless propagators, which are completely known at four loops \cite{Baikov:2010hf,Lee:2011jt},
or some specific single-scale problems (see e.g.\ \cite{Vladimirov:1979zm} for a two-loop example relevant for the present work).
In the present paper, we add one massive two-loop example to the pool of general solutions, 
be it as ingredient for optimizing higher-order calculations, or as a benchmark for alternative evaluation methods.

To motivate the specific choice of masses that we study here, let us briefly recall the \kallen function, perhaps mainly known
from its role in kinematics.
The completely symmetric \kallen function \cite{Kallen:1964lxa} of 3 arguments 
\ba \la{eq:defKallen}
\lambda(x^2,y^2,z^2) &\equiv& x^4+y^4+z^4-2(x^2y^2+y^2z^2+z^2x^2) \\ \la{eq:defKallen2}
&=& (x+y+z)(x-y-z)(y-z-x)(z-x-y)
\ea
is known to play a prominent role in multi-particle scattering processes, due to its relevance in relativistic particle kinematics and phase space distributions \cite{Byckling:1971vca}.
It is sometimes also called triangle function, since in geometry Heron's formula asserts that $\sqrt{-\lambda(x^2,y^2,z^2)}$ measures (four times) the area of a triangle with side lengths $x$, $y$ and $z$.
Here, we will encounter this function multiple times, in quite a different -- but in view of \se\ref{se:applications} maybe not completely unrelated -- context.

We note that in IBP reductions of Feynman integrals to their respective sets of master integrals the \kallen function appears in a number of explicit integral reduction relations, as a function of squared external four-momenta or Mandelstam variables and squared masses.
For concrete examples, we refer to \cite{Berends:1996gs} as well as the seminal paper \cite{Tarasov:1997kx}, where the integral reduction problem for the case of two-loop two-point functions had been solved in complete generality, in terms of a recursive algorithm.
It might therefore be natural to ask whether the integral reduction problem is affected by zeros of the \kallen function.

In order to get a first insight by studying the cleanest possible setting, let us focus on Feynman integrals where all external momenta vanish, so-called vacuum integrals. This leaves only the internal particles' masses as energy scales. Furthermore, since we need three arguments for the \kallen function, we restrict ourselves to the two-loop level, where the generic scalar vacuum integral 
(we use Euclidean notation, and work in dimensional regularization \cite{Bollini:1972ui,tHooft:1972tcz,Ashmore:1972uj,Cicuta:1972jf} with $d$ spacetime dimensions)
\ba
\la{eq:defB1}
\intB^\nabc(d;m_1^2,m_2^2,m_3^2) &\equiv&
\intd{p} \intd{q}
\frac{1}{[m_1^2+{p}^2]^{\nu_1}\,[m_2^2+{q}^2]^{\nu_2}\,[m_3^2+({p}-{q})^2]^{\nu_3}} 
\ea
has three different propagators and hence allows for three different mass scales. Zeros of the \kallen function are then given whenever one of the masses is the sum of two others, 
as can be seen from its factorized form \eq\nr{eq:defKallen2}, or from its geometric interpretation: when one triangle side is equal to the two others, the triangle's area is obviously zero. Without loss of generality, we choose all masses non-negative and the 'large' mass to be $m_3$, such that for the purpose of this paper 
\ba\la{eq:massRel}
\lambda(m_1^2,m_2^2,m_3^2) \;=\; 0 \quad\Leftrightarrow\quad \fbox{$m_3=m_1+m_2$} \;.
\ea
To avoid ambiguities, let us slightly change notation when referring to two-loop vacuum integrals that imply this condition and denote masses as subscripts as
\ba
\la{eq:defB2}
\intB^\nabc_\mabc(d) &\equiv&
\intB^\nabc(d;m_1^2,m_2^2,m_3^2) \Big|_{m_3=m_1+m_2} \;.
\ea
For a graphical representation, see \fig\ref{fig:B123}.
Even though we have fixed $m_3\equiv m_1+m_2$, we will keep using $m_3$ as an index of the functions $\intB$, 
since this at times makes symmetries more explicit.
We further note that the linear mass relation of \eq\nr{eq:massRel} renders our configuration 
``doubly special", since the vacuum diagrams can not only be considered as two-point sunset functions 
at vanishing external momentum $q$, but also because in this case one is at the sunset's 
pseudo-threshold $q^2=(m_1+m_2-m_3)^2=0$.

\begin{figure}
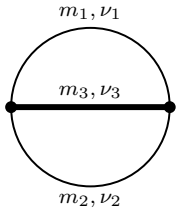

\setlength{\unitlength}{1.5pt}
\ba \nonumber
\sunB{$m_1,\nu_1$}{$m_2,\nu_2$}{$m_3,\nu_3$} 
\ea
\caption{\la{fig:B123}The main object of the paper, the massive two-loop vacuum integral as defined in \eq\nr{eq:defB2}. 
In our graphical representation, each line corresponds to a massive scalar propagator $1/[m^2+p^2]^\nu$, with respective mass and power as indicated.  
We use a thick line to mark the special propagator mass $m_3=m_1+m_2$.}
\end{figure}

In addition to this rather formal motivation, we note that linear relations between propagator masses such 
as in \eq\nr{eq:massRel} play a prominent role in QCD thermodynamics, where a core phenomenological problem 
concerns describing the equilibrium properties of a hot plasma of gauge bosons,
codified by weak-coupling expansions \cite{Braaten:1995jr,Kajantie:2002wa}
within finite-temperature field theory (for a modern introduction to this field, see e.g.\ \cite{Laine:2016hma}). 
In this setting, the role of bare propagator masses is played
by the temporal components of their momentum four-vectors, such that energy-momentum conservation at interaction vertices leads 
to such linear ``mass'' relations. In fact, our initial interest in pursuing the project reported upon here stems from concrete perturbative
expansions for the QCD pressure at higher loop orders \cite{Navarrete:2022rbt}. Temperature will not play a role at all in the present paper, however, such that its results might
be useful for practitioners of ordinary (i.e.\ zero-temperature) perturbative quantum field theory.

The structure of the remainder of the paper is as follows. In \se\ref{se:A}, we recall an IBP relation for our main object of study, 
the integral of \eq\nr{eq:defB2}, together with a well-known special-mass case. We then prepare grounds towards a general 
solution of the IBP recurrence in \se\ref{se:closed}, by first obtaining closed forms for special cases when two of the integral's 
propagator indices coincide, whose analytic form then leads us to a general conjecture for the mass-dependence of the integral, 
which we manage to prove based on the IBP relation. Section~\ref{se:twoStep} then employs an additional recurrence relation 
that allows to map the index-dependence of the integral to a purely combinatorial problem, whose solution can finally be obtained 
by comparing with the special cases considered before, thus leading to a general factorization formula. Equipped with the general 
solution, in \se\ref{se:applications} we recall the relation between two-loop vacuum and one-loop triangle integrals, and apply our 
result to the latter case, before summarizing and concluding in \se\ref{se:summary}. To not interrupt the flow of 
argument in the main text, we have relegated two proofs, a brief exposure of one of the recurrence 
relations relevant in the main text as well as an analytic treatment of one interesting special-mass case to the appendices.

%
\section{Massive two-loop vacuum integrals: IBP recursions}
\la{se:A}

For the massive two-loop scalar vacuum integrals $\intB^\nabc(d;m_1^2,m_2^2,m_3^2)$ that were defined in \eq\nr{eq:defB1}, general results in terms of Appell's hypergeometric function $F_4$ can be 
found\footnote{\la{foot:1}Using \eq(2.7) of \cite{Davydychev:1992mt} to convert to our Euclidean conventions, 
we have $I(d;\nu_1,\nu_2,\nu_3 |\,m_1,m_2,m_3) = (-1)^{\Nu+1}\,\pi^d\,L^{-2d}\,\intB_\mabc^\nabc(d)$, with factor
$L=(4\pi)^{-1/2}$ due to our integral measure \eq\nr{eq:defB1} (cf.\ footnote \ref{ft:L}).} in \eq(4.3) of \cite{Davydychev:1992mt}.
As has been mentioned above, in the case of interest to us here, the masses obey a ``mass conservation rule'' at each vertex, $m_3=m_1+m_2$. 
For this special case $\intB^\nabc_\mabc(d)$, cf.\ \eq\nr{eq:defB2}, the general results of \cite{Davydychev:1992mt} can be simplified. 
Instead of dealing with the rather complicated $F_4$ functions (we revisit them in \app\ref{se:112}), however, 
we choose to attack the problem from a different angle here and exploit integration-by-parts (IBP) identities, specialized to our specific mass restriction from the outset. 

IBP identities \cite{Chetyrkin:1981qh,Tkachov:1981wb} provide linear relations between dimensionally-regulated 
Feynman integrals, and can be used to systematically reduce the set of positive propagator powers $\nu_i\in \setN$ 
towards lower values \cite{Tarasov:1997kx,Laporta:2000dsw}. 
In the case at hand, it turns out that the IBP identities provide recursive relations that reduce 
all $\nu_i$ to one or zero, as explained in the subsequent sections. 

As indicated by the notation \eq\nr{eq:defB2}, we implicitly assume $m_3=m_1+m_2$ in all equations below, and will often denote 
the sum of propagator powers as $\Nu\equiv\nu_1+\nu_2+\nu_3$.
A number of special mass-cases of our integral are known analytically in $d$ dimensions, such as
\ba
\intB_{0,0,0}^\nabc(d) &=& 0 \mbox{~~(massless tadpoles vanish in dimensional regularization)},\\
\tB^\nabc(d) &\propto& \mbox{~product of Gamma functions \cite{Vladimirov:1979zm}, see \eq\nr{eq:B0mmRes} below},\\
\hB^\nabc(d) &\propto& \mbox{~hypergeometric function $\Fdc(\dots|1)$ \cite{Davydychev:1992mt}, see \app\ref{se:112} below}.
\ea
The case when two masses vanish is also known analytically, of course, 
but this is not relevant for our purposes here since it lies outside our class of integrals that fulfill $m_3=m_1+m_2$
and rather corresponds to the general integrals of \eq\nr{eq:defB1} as $\intB^\nabc(d;0,0,m^2)$.

%
\subsection{General mass case $[m_1,m_2,m_1+m_2]$}
\la{se:recu}

Equation~(95) of (the journal version of) \cite{Tarasov:1997kx} (with a small correction\footnote{We thank 
Oleg Tarasov for confirming this misprint.}: $d\rightarrow d+2$) provides a recursive 2-loop sunset factorization
\ba \la{eq:T92}
\intB_\mabc^\nabc(d) &=&
\frac{-1}{2(d+3-2\Nu)m_1m_2m_3}
\,\Big\{\nonumber\\
&& +\big[ (m_1 (d + 2 - \Nu) + m_2 \nu_3 - m_3 \nu_2) \big] \,\subA \nonumber\\
&& +\big[ (m_1 \nu_3 + m_2 (d + 2 - \Nu) - m_3 \nu_1) \big] \,\subB \nonumber\\
&& +\big[ (m_1 \nu_2 + m_2 \nu_1 - m_3 (d + 2 - \Nu)) \big] \,\subC \,\Big\}\,\intB_\mabc^\nabc(d) \;,
\ea
which can be used until one of the $\nu_i$ is zero.
We have written the relation in terms of lowering operators, which act on the integral's indices as 
\ba\la{eq:loweringOp}
\subA\,\intB_\mabc^\nabc(d) &\equiv& \intB_\mabc^{\nu_1-1,\nu_2,\nu_3}(d) \;,
\ea
plus similar relations for the remaining indices.
We will omit the mass indices of the integrals $\intB$ for brevity below, when no confusion can arise.
Boundary conditions for the 2-loop recursion \eq\nr{eq:T92} are products of 1-loop tadpoles
which either vanish in dimensional regularization if they are scale-free
\ba \la{eq:bc1}
\intB^{0,0,0}(d) &=& \intB^{\nu_1,0,0}(d) = \intB^{0,\nu_2,0}(d) = \intB^{0,0,\nu_3}(d) = 0 \;,
\ea
or which can be trivially reduced to a unique representative or master integral (for example, with unit propagator powers) 
by polynomial prefactors $h_{a,n}(d)$ as
\ba
\la{eq:B120}
\intB^{\nu_1,\nu_2,0}(d) &=& 
\frac{h_{\nu_1,0}(d)}{m_1^{2\nu_1-2}}\,
\frac{h_{\nu_2,0}(d)}{m_2^{2\nu_2-2}}\,
\intB^{1,1,0}(d)
\;,\\
\la{eq:bc3}
\intB^{\nu_1,0,\nu_3}(d) &=& 
\frac{h_{\nu_1,0}(d)}{m_1^{2\nu_1-2}}\,
\frac{h_{\nu_3,0}(d)}{m_3^{2\nu_3-2}}\,
\intB^{1,0,1}(d) \;,\\
\la{eq:bc4}
\intB^{0,\nu_2,\nu_3}(d) &=& 
\frac{h_{\nu_2,0}(d)}{m_2^{2\nu_2-2}}\,
\frac{h_{\nu_3,0}(d)}{m_3^{2\nu_3-2}}\,
\intB^{0,1,1}(d) \;,
\ea
where the prefactors are defined by one-loop tadpole reduction\footnote{\la{ft:L}The factor $L$ is pure convention, 
the value given here being a consequence of our integral measure \eq\nr{eq:defB1}. It could be defined as $L=1$ by changing the 
measure to $\int\!\frac{{\rm d}^dp}{\pi^{d/2}}$, but we keep it here for ease of comparison with the literature.}
\ba
&&\int\!\frac{{\rm d}^{d-2n}p}{(2\pi)^{d-2n}}\,\frac1{[1+p^2]^a} = L^{-2n}\,h_{a,n}(d) \int\!\frac{{\rm d}^{d}p}{(2\pi)^d}\,\frac1{[1+p^2]}
\;,\\ \la{eq:defh}
&&\mbox{with~~}h_{a,n}(d)\equiv\frac{\Gamma(a+n-d/2)}{\Gamma(a)\,\Gamma(1-d/2)}
\mbox{~~and~~}L=(4\pi)^{-\frac12}
\;.
\ea
In our normalization, the three master integrals read
\ba \la{eq:masters}
\intB^{1,1,0}(d) &=& \frac{L^{2d}\,\Gamma^2(1-d/2)}{m_1^{2-d}m_2^{2-d}} \;,
\intB^{1,0,1}(d) = \frac{L^{2d}\,\Gamma^2(1-d/2)}{m_1^{2-d}m_3^{2-d}} \;,
\intB^{0,1,1}(d) = \frac{L^{2d}\,\Gamma^2(1-d/2)}{m_2^{2-d}m_3^{2-d}} \;.
\ea

With the trivially factorized cases \eqs\nr{eq:bc1}-\nr{eq:bc4} out of the way, we will henceforth implicitly assume positive integer 
indices $\nu_i$.
For the special case of equal indices $\nu_1=\nu_2=\nu_3\equiv\nu$, \eq\nr{eq:T92} reads
\ba
\intB_\mabc^{\nu,\nu,\nu}(d) &=& \frac{-(d+2-4\nu)}{2(d+3-6\nu)}\, \Big\{ \frac{\subA}{m_2m_3} +\frac{\subB}{m_1m_3} -\frac{\subC}{m_1m_2} \Big\}\,\intB_\mabc^{\nu,\nu,\nu}(d) \;,
\ea
which at $\nu=1$ gives the integral in a factorized form (cf.\ \fig\ref{fig:B111}; again omitting mass indices of $\intB$), 
\ba \la{eq:111}
\intB^{1,1,1}(d) &=& \frac{-(d-2)}{2(d-3)}\, 
\Big\{ \frac{\intB^{0,1,1}(d)}{m_2m_3} +\frac{\intB^{1,0,1}(d)}{m_1m_3} -\frac{\intB^{1,1,0}(d)}{m_1m_2} \Big\} \;.
\ea
Note that for the general-mass case $\intB^{1,1,1}$ would have been a master integral \cite{Davydychev:1992mt}, 
the reduction to one-loop tadpoles here being due to our mass relation \eq\nr{eq:massRel}.

\begin{figure}
\setlength{\unitlength}{1.3pt}
\ba \nonumber
\sunB{$m_1,1$}{$m_2,1$}{$m_3,1$} &=& \frac{(d-2)}{2(d-3)}\,\Bigg[
\frac1{m_1\,m_2}\,\tadA{$m_1,1$}{$m_2,1$}
-\frac1{m_1\,m_3}\,\tadB{$m_1,1$}{$m_3,1$}
-\frac1{m_2\,m_3}\,\tadB{$m_2,1$}{$m_3,1$}
\;\Bigg]
\ea
\caption{\la{fig:B111}Visualization of the factorization of a two-loop integral with unit propagator powers into a basis of one-loop tadpoles, 
as given in \eq\nr{eq:111} (see also \eq\nr{eq:B111}). The graphical notation is as in \fig\ref{fig:B123}.}
\end{figure}

%
\subsection{Special mass case $[0,m,m]$}
\la{se:B0mm}

For the special case with masses $[0,m,m]$, \eq(96) of (the journal version of) \cite{Tarasov:1997kx} gives a recursive 2-loop sunset factorization 
\ba \la{eq:T93}
\tB^\nabc(d) &=& a_{\nu_1-1}\tB^{\nu_1-1,\nu_2,\nu_3}(d) 
\ea
with rational coefficient function
\ba
a_{\nu_1} &=& -\frac{(d-2\nu_1-2\nu_2)(d-2\nu_1-2\nu_3)(d-\nu_1-\nu_2-\nu_3)}{2m^2(d-2-2\nu_1)(d-1-2\nu_1-\nu_2-\nu_3)(d-2\nu_1-\nu_2-\nu_3)} \;.
\ea
The boundary condition is $\tB^{0,\nu_2,\nu_3}=\frac{h_{\nu_2,0}(d)h_{\nu_3,0}(d)}{m^{2\nu_2+2\nu_3-4}}\,\tB^{0,1,1}$, with master integral $\tB^{0,1,1}=\frac{L^{2d}\,\Gamma^2(1-d/2)}{m^{4-2d}}$.
The integral can therefore be solved easily in closed form by iteration to $\nu_1=0$ (cf.\ \eq\nr{eq:C36}), 
followed by tadpole reduction to $\nu_2=\nu_3=1$:
\ba\la{eq:B0mmRes}
\tB^\nabc &=&
\Big(\prod_{j=0}^{\nu_1-1} a_j\Big)\,\tB^{0,\nu_2,\nu_3} 
\;=\; \Big(\prod_{j=0}^{\nu_1-1} a_j\Big)\,\frac{h_{\nu_2,0}(d)h_{\nu_3,0}(d)\tB^{0,1,1}}{m^{2\nu_2+2\nu_3-4}} 
\;=\; \bzmm\nabc\,\frac{\tB^{0,1,1}}{(m^2)^{\Nu-2}}  \qquad
\\
\mbox{with}&&\bzmm\nabc\;=\; \la{eq:B0mmAlt}
\frac{(-1)^{\nu_1}\Gamma(\nu_1+\nu_2-d/2)\Gamma(\nu_1+\nu_3-d/2)\Gamma(\Nu-d)}
{\Gamma(\nu_2)\Gamma(\nu_3)\Gamma(1-d/2)\Gamma(1-d/2+\nu_1)\Gamma(\nu_1+\Nu-d)} \\
&&\hphantom{\bzmm\nabc}\;=\; \la{eq:B0mm}
\frac{\Gamma(\nu_1+\nu_2-d/2)\Gamma(\nu_1+\nu_3-d/2)\Gamma(d/2-\nu_1)\Gamma(\Nu-d)}
{\Gamma(\nu_2)\Gamma(\nu_3)\Gamma^2(1-d/2)\Gamma(d/2)\Gamma(\nu_1+\Nu-d)} \;,
\ea
where in the last step we have used the Euler reflection formula to identify 
$\frac{(-1)^{\nu_1}}{\Gamma(1-d/2+\nu_1)}=\frac{\Gamma(d/2-\nu_1)}{\Gamma(d/2)\Gamma(1-d/2)}$ 
on the integers $\nu_1$.
This simple result has been known for a long time already, the original reference probably being \eq(A.6) of \cite{Vladimirov:1979zm}.

%
\section{Massive two-loop vacuum integrals: Towards closed-form results}
\la{se:closed}

We would now like to solve the 3-parameter recurrence \eq\nr{eq:T92} for the integral $\intB^\nabc_\mabc(d)$ in closed form as well. Unfortunately, we have not succeeded in doing so directly from the recurrence relation. We did succeed, however, in deriving a closed-form (i.e.\ non-recursive) solution by first looking at simpler special cases where two indices $\nu_i$ coincide (see \secs\ref{se:B11n} and \ref{se:Baac} below), 
then extracting the polynomial dependence on masses to all orders for general indices (cf.\ \secs\ref{se:conj} and \ref{se:conjProof}), extracting the polynomial dependence on the dimension $d$\ by solving an additional ($d$\/-dependent) recurrence in terms of a purely combinatorial function (\se\ref{se:4.1}), the latter of which could finally be determined by comparing with the special-index cases (we explain this step in \se\ref{se:4.2}). Our final result is given by \eqs\nr{eq:newconj} and \nr{eq:kallen3}.

%
\subsection{Special index cases $\intB^{1,1,\nu}$, $\intB^{1,\nu,1}$, $\intB^{\nu,1,1}$}
\la{se:B11n}

For the special case $\intB^{1,1,\nu}(d)$, one step of the IBP recursion \eq\nr{eq:T92} has the simple structure
\ba\la{eq:inhom}
\intB^{1,1,\nu}(d) &=& a_{\nu-1}\,\intB^{1,1,\nu-1}(d) +b_\nu \;,
\ea
where the coefficient functions are given by
\ba
a_\nu &=& \frac{m_1+m_2+(\nu+1-d)m_3}{4m_1m_2m_3(\nu+\frac{3-d}2)} = \frac{(2-d+\nu)}{4m_1m_2\,(\frac{3-d}2+\nu)} \;,\\
\la{eq:fn}
b_\nu &=& \frac{[\nu\,m_1+(d-\nu)m_2-m_3]}{4(\nu+\frac{1-d}2)m_1m_2m_3}\,\intB^{1,0,\nu}(d)
+\Big(\begin{array}{c}\nu_1\leftrightarrow\nu_2\\m_1\leftrightarrow m_2\end{array}\Big) \;.
\ea
Iterating \eq\nr{eq:inhom} until hitting the boundary value at $\nu=0$, one obtains (cf.\ \eqs\nr{eq:C35}, \nr{eq:C36})
\ba\la{eq:inhomSoln}
\intB^{1,1,\nu}(d) &=&  \frac{\po{2-d}{\nu}}{\po{\frac{3-d}2}{\nu}}\,\frac{\intB^{1,1,0}(d)}{(4m_1m_2)^\nu} 
+ \sum_{j=1}^{\nu} \frac{\po{2-d+j}{\nu-j}}{\po{\frac{3-d}2+j}{\nu-j}}\,\frac{b_j}{(4m_1m_2)^{\nu-j}} \;,
\ea
where 
\ba \la{eq:defPoch}
\po{a}{\nu} &\equiv& \frac{\Gamma(a+\nu)}{\Gamma(a)}
\ea
are Pochhammer symbols. 
The second term of \eq\nr{eq:inhomSoln} can be simplified. 
Considering only the first part of $b_\nu$ as given in \eq\nr{eq:fn} (the second part then follows trivially from the indicated $1\leftrightarrow2$ replacement), 
we employ the tadpole reduction \eq\nr{eq:bc3} to reduce $\intB^{1,0,\nu}(d)$ to the master integral $\intB^{1,0,1}(d)$ and change $m_2$ into $m_3-m_1$ where convenient,
such that the $j$\/-sum reads 
\ba\la{eq:417}
\sum_{j=1}^{\nu} \frac{\po{2-d+j}{\nu-j}}{\po{\frac{3-d}2+j}{\nu-j}}\,
\frac{\po{1-\frac{d}2}{j-1}[(d-j-1)+(2j-d)\frac{m_1}{m_3}]}{\Gamma(j)\,(j+\frac{1-d}2)\,4^{\nu-j+1}\,(\frac{m_1}{m_3})^{1-j}\,(1-\frac{m_1}{m_3})^{1-j+\nu}}\, 
\frac{\intB^{1,0,1}(d)}{(m_1\,m_3)^\nu} \;.
\ea
Expanding the summand and collecting powers of the mass ratio, this is equivalent to (see \app\ref{se:418proof})
\ba\la{eq:418}
-\frac{\intB^{1,0,1}(d)}{(m_1\,m_3)^\nu} 
\sum_{j=0}^{\nu-1} \frac{2\po{1-\frac{d}2}{\nu+1}}{j!\,\po{2-d+\nu+j}{\nu+1-j}}\, \Big(\frac{m_1}{m_3}\Big)^j\;.
\ea
Collecting, and recording the cases $\intB^{1,\nu,1}$ and $\intB^{\nu,1,1}$ as well, we finally obtain
\ba \la{eq:B11n}
\frac{\intB^{1,1,\nu}(d)}{\beta^{\nu,1,1}}
&=& \frac{\intB^{1,1,0}(d)}{(-m_1m_2)^\nu}
-\sum_{j=0}^{\nu-1}
\frac{\po{2-d+\nu}{j}}{j!}
\Big\{
\frac{\intB^{1,0,1}(d)}{(-m_1m_3)^\nu}\Big(\frac{m_1}{m_3}\Big)^j
+\frac{\intB^{0,1,1}(d)}{(-m_2m_3)^\nu}\Big(\frac{m_2}{m_3}\Big)^j
\Big\} \;,\qquad\\
\frac{\intB^{1,\nu,1}(d)}{\beta^{\nu,1,1}}
&=& \frac{\intB^{1,0,1}(d)}{(m_1m_3)^\nu}
-\sum_{j=0}^{\nu-1}
\frac{\po{2-d+\nu}{j}}{j!}
\Big\{
\frac{\intB^{1,1,0}(d)}{(m_1m_2)^\nu}\Big(-\frac{m_1}{m_2}\Big)^j
+\frac{\intB^{0,1,1}(d)}{(-m_3m_2)^\nu}\Big(\frac{m_3}{m_2}\Big)^j
\Big\} \;,\qquad\\
\frac{\intB^{\nu,1,1}(d)}{\beta^{\nu,1,1}}
&=& \frac{\intB^{0,1,1}(d)}{(m_2m_3)^\nu}
-\sum_{j=0}^{\nu-1}
\frac{\po{2-d+\nu}{j}}{j!}
\Big\{
\frac{\intB^{1,1,0}(d)}{(m_2m_1)^\nu}\Big(-\frac{m_2}{m_1}\Big)^j
+\frac{\intB^{1,0,1}(d)}{(-m_3m_1)^\nu}\Big(\frac{m_3}{m_1}\Big)^j
\Big\} \;, \la{eq:Bn11}
\ea
where the normalization factor $\beta^{\nu,1,1}$ is a special case of \eq\nr{eq:B0mm},
\ba
\beta^{\nu,1,1}&=&\frac{\po{1-\tfrac{d}2}{\nu}\,\po{1-\tfrac{d}2}{\nu}}{\po{\tfrac{d}2-\nu}{\nu}\,\po{2-d+\nu}{\nu}} 
\;=\;\frac{(-1)^\nu \po{1-\frac{d}2}{\nu}}{\po{2-d+\nu}{\nu}} 
\;=\;\frac{1}{(-4)^\nu}\,\frac{\po{2-d}{\nu}}{\po{\frac{3-d}2}{\nu}} \;.
\ea

As a check, for $\nu=1$ all three equations reduce to \eq\nr{eq:111} (since $\beta^{1,1,1}=-\frac{(d-2)}{2(d-3)}$).
Alternatively, the results \eqs\nr{eq:B11n}-\nr{eq:Bn11} can be derived from \eqs(7) and (15) of (the journal version of) \cite{Davydychev:1995mq}.

%
\subsection{Special index cases $\intB^{\ka,\ka,\nu}$, $\intB^{\ka,\nu,\ka}$, $\intB^{\nu,\ka,\ka}$}
\la{se:Baac}

Having $\intB^{1,1,\nu}(d)$ (and permutations) at hand, one can go one step further and derive closed forms for the 
integrals $\intB^{\ka,\ka,\nu_3}(d)$ (and permutations).
For example, repeatedly using the second of \eqs(94) of \cite{Tarasov:1997kx} (journal version; converted to our notation)
\ba \la{eq:T94b}
(d-2)\nu_1\nu_2\,\upA\,\upB\,\intB^\nabc(d) &=& 
\Big\{(d-2-2\nu_3)+2m_3^2\nu_3\,\upC\Big\}\,\subDD\,\intB^\nabc(d) \;,
\ea
which employs index-raising operators that act in analogy to \eq\nr{eq:loweringOp} as e.g.\ $\upC \intB^\nabc=\intB^{\nu_1,\nu_2,\nu_3+1}$, as well as a dimension-shift operator $\subDD$ that acts on our integrals 
according to\footnote{The normalization factor $L^4=\frac1{16\pi^2}$ is a consequence of our integral measure, see also footnote \ref{ft:L} above.}
\ba\la{eq:subDD}
\subDD\,\intB^\nabc(d)&\equiv&L^4\, \intB^\nabc(d-2) \;,
\ea
we obtain
\ba
(\upA\upB)^N \intB^{1,1,\nu}(d) &=&
\la{eq:Baac}
\sum_{k=0}^N \frac{\po{\nu}{k} \po{1+N-\tfrac{d}2}{\nu}}{N!\,k!\,(N-k)!\,\po{1-\tfrac{d}2}{k+\nu}} 
(-m_3^2\,\upC)^k (\subDD)^N \intB^{1,1,\nu}(d) \;.
\ea
This is actually the first non-trivial recurrence that we solve in this paper, and it is instructive to expose its guts, which we have done in \app\ref{app:guts}. It turns out that at the heart of the IBP relation \eq\nr{eq:T94b} lies a two-dimensional linear homogeneous recurrence relation with variable coefficients (cf.\ \eq\nr{eq:guts}), which despite its close resemblance to the Stirling recurrence admits a closed-form solution.

Once the expression \eq\nr{eq:Baac} has been derived, it can alternatively be proven directly by induction over $N$:  
at $N=0$, \eq\nr{eq:Baac} reduces to $\intB^{1,1,\nu}(d)=\intB^{1,1,\nu}(d)$, which is trivially valid. 
Assume now that it holds for one non-negative integer $N$. Then, writing $(\upA\upB)^{N+1}\intB^{1,1,\nu}(d)=\upA\upB\intB^{1+N,1+N,\nu}(d)$, using \eq\nr{eq:T94b} at $\nu_1=\nu_2=1+N$ on the right-hand side, using the assumption for both resulting integrals $\intB^{1+N,1+N,\nu}(d-2)$ and $\intB^{1+N,1+N,\nu+1}(d-2)$, and rearranging terms, \eq\nr{eq:Baac} is seen to hold at $N+1$ as well, which completes the proof.

The integrals on the right-hand side of \eq\nr{eq:Baac} can now be written in terms of master integrals via \eq\nr{eq:B11n}, after which those can be shifted back to a common dimension $d$ using \eq\nr{eq:masters}, i.e.\ via 
\ba\la{eq:tad2}
\frac{\subDD\intB^{1,1,0}(d)}{\intB^{1,1,0}(d)}&=& 
\frac{L^4\,\intB^{1,1,0}(d-2)}{\intB^{1,1,0}(d)}
=\frac{(1-\tfrac{d}2)^2}{m_1^2m_2^2}
\ea 
and permutations; the resulting expression is (as above, we keep $m_3$ to emphasize symmetry)
\ba \la{eq:BaacM}
\intB^{\ka,\ka,\nu}(d) &=& \sum_{k=0}^{\ka-1} \frac{(-1)^k \po{\nu}{k} \po{1-\tfrac{d}2}{\ka+\nu-1} \po{1+\nu+k-\tfrac{d}2}{\ka-1}}{(\ka-1)!\,k!\,(\ka-1-k)!\,\po{2\ka+\nu+k-d}{\nu+k}}
\Bigg\{
\frac{\intB^{1,1,0}(d)}{(m_1m_2)^{2\ka+\nu-2}}\,\frac{(m_1+m_2)^{2k}}{(m_1m_2)^k}
\nonumber\\
&-&\sum_{j=0}^{\nu+k-1} \frac{\po{2\ka\!+\!\nu\!+\!k\!-\!d}{j}}{j!}
\Big[
\frac{\intB^{1,0,1}(d)}{(m_1m_3)^{2\ka+\nu-2}}\,\Big(\frac{m_1}{m_3}\Big)^{j-k}
+\frac{\intB^{0,1,1}(d)}{(m_2m_3)^{2\ka+\nu-2}}\,\Big(\frac{m_2}{m_3}\Big)^{j-k}
\Big]
\Bigg\}
\;.\qquad
\ea
The corresponding results for the permutations read (note a certain symmetry in $\{m_1,m_2,-m_3\}$)
\ba \la{eq:BacaM}
\intB^{\ka,\nu,\ka}(d) &=& \sum_{k=0}^{\ka-1} \frac{(-1)^k \po{\nu}{k} \po{1-\tfrac{d}2}{\ka+\nu-1} \po{1+\nu+k-\tfrac{d}2}{\ka-1}}{(\ka-1)!\,k!\,(\ka-1-k)!\,\po{2\ka+\nu+k-d}{\nu+k}}
\Bigg\{
\frac{\intB^{1,0,1}(d)}{(-m_1m_3)^{2\ka+\nu-2}}\,\frac{(m_3-m_1)^{2k}}{(-m_1m_3)^k}
\nonumber\\&-&
\sum_{j=0}^{\nu+k-1} \frac{\po{2\ka\!+\!\nu\!+\!k\!-\!d}{j}}{j!}
\Big[
\frac{\intB^{1,1,0}(d)}{(-m_1m_2)^{2\ka+\nu-2}}\,\Big(\!\!-\!\frac{m_1}{m_2}\Big)^{j-k}
\!\!\!+\!\frac{\intB^{0,1,1}(d)}{(m_2m_3)^{2\ka+\nu-2}}\,\Big(\frac{m_3}{m_2}\Big)^{j-k}
\Big]
\Bigg\}
,\qquad
\\ \la{eq:BcaaM}
\intB^{\nu,\ka,\ka}(d) &=& \sum_{k=0}^{\ka-1} \frac{(-1)^k \po{\nu}{k} \po{1-\tfrac{d}2}{\ka+\nu-1} \po{1+\nu+k-\tfrac{d}2}{\ka-1}}{(\ka-1)!\,k!\,(\ka-1-k)!\,\po{2\ka+\nu+k-d}{\nu+k}}
\Bigg\{
\frac{\intB^{0,1,1}(d)}{(-m_2m_3)^{2\ka+\nu-2}}\,\frac{(m_3-m_2)^{2k}}{(-m_2m_3)^k}
\nonumber\\&-&
\sum_{j=0}^{\nu+k-1} \frac{\po{2\ka\!+\!\nu\!+\!k\!-\!d}{j}}{j!}
\Big[
\frac{\intB^{1,1,0}(d)}{(-m_1m_2)^{2\ka+\nu-2}}\,\Big(\!\!-\!\frac{m_2}{m_1}\Big)^{j-k}
\!\!\!+\!\frac{\intB^{1,0,1}(d)}{(m_1m_3)^{2\ka+\nu-2}}\,\Big(\frac{m_3}{m_1}\Big)^{j-k}
\Big]
\Bigg\}
.\qquad
\ea

%
\subsection{Conjectured general structure for $\intB^\nabc$}
\la{se:conj}

While we now have the general solution for the special-index cases $\intB^{\ka,\ka,\nu}(d)$ (plus index permutations) available, 
we do not seem to be able to generalize this further to $\intB^{\nabc}(d)$, in order to obtain a general solution of the three-parameter recursion \eq\nr{eq:T92}. 

Inspecting the solutions \eq\nr{eq:BaacM}-\nr{eq:BcaaM}, though, we notice a pattern: the denominator corresponding to a master integral with vanishing index $\nu_i$ only contains powers of masses that do {\em not} involve $m_i$. Furthermore, we note the similarity of prefactors of three the distinct master integrals, including some apparent systematics in the signs that seems to be linked to the presence of $m_3$. These patterns can then be verified by examining a number of fixed-index examples, which in turn are easily generated by an implementation of the recursion \eq\nr{eq:T92}, followed by tadpole reduction to masters and a convenient rewriting of mass factors using \eq\nr{eq:massRel}. 
Some examples are (as always, $m_3=m_1+m_2$ is implied)
\ba
\la{eq:B111}
\intB^{1,1,1}(d) &=& 
\frac{(d-2)}{2(d-3)} \Bigg\{
-\frac{\intB^{0,1,1}(d)}{m_2 m_3}
-\frac{\intB^{1,0,1}(d)}{m_1 m_3}
+\frac{\intB^{1,1,0}(d)}{m_1 m_2}
\Bigg\}
\\\intB^{2,1,1}(d) &=& 
\frac{(d-2)}{4 (d-5)} \Bigg\{
\frac{\intB^{0,1,1}(d)}{m_2^2 m_3^2}
+\Big[(d-4) \tfrac{m_3}{m_1}-1\Big] \frac{\intB^{1,0,1}(d)}{m_1^2 m_3^2}
-\Big[(d-4)\tfrac{m_2}{m_1}+1\Big] \frac{\intB^{1,1,0}(d)}{m_1^2 m_2^2}
\Bigg\}
\\\intB^{1,2,1}(d) &=& 
\frac{(d-2)}{4 (d-5)} \Bigg\{
\Big[(d-4)\tfrac{m_3}{m_2}-1\Big] \frac{\intB^{0,1,1}(d)}{m_2^2 m_3^2}
+\frac{\intB^{1,0,1}(d)}{m_1^2 m_3^2}
-\Big[(d-4) \tfrac{m_1}{m_2}+1\Big] \frac{\intB^{1,1,0}(d)}{m_1^2 m_2^2}
\Bigg\}
\\\intB^{1,1,2}(d) &=& 
\frac{(d-2)}{4 (d-5)} \Bigg\{
\Big[(d-4) \tfrac{m_2}{m_3}-1\Big] \frac{\intB^{0,1,1}(d)}{m_2^2 m_3^2}
+\Big[(d-4)\tfrac{m_1}{m_3}-1\Big] \frac{\intB^{1,0,1}(d)}{m_1^2 m_3^2}
+\frac{\intB^{1,1,0}(d)}{m_1^2 m_2^2}
\Bigg\}
\\\intB^{1,2,3}(d) &=& \la{eq:exampleB123}
\frac{(d-4) (d-2)}{32 (d-9) (d-7)} \Bigg\{
-\Big[2 (d^2-11 d+27)+(d-7) (d-6) \big(\tfrac{m_2}{m_3}\big)^2 -(d-8) (d-6) (d-3) \tfrac{m_2}{m_3}
\nonumber\\&&
-2 (d-6)\tfrac{m_3}{m_2}\Big] \frac{\intB^{0,1,1}(d)}{m_2^4 m_3^4}
+\Big[(d-7) (d-6) \big(\tfrac{m_1}{m_3}\big)^2 -4 (d-6)\tfrac{m_1}{m_3} +6\Big] \frac{\intB^{1,0,1}(d)}{m_1^4 m_3^4}
\nonumber\\&&
-2\Big[(d-6) \tfrac{m_1}{m_2}+3\Big] \frac{\intB^{1,1,0}(d)}{m_1^4 m_2^4} 
\Bigg\}
\;.
\ea
Looking at a number of examples like those above, we are led to conjecture that (cf.\ \fig\ref{fig:conj})
\ba \la{eq:newconj}
\intB^\nabc_\mabc(d) &=& 
\frac{\intB^{1,1,0}_\mabc(d)}{(m_1m_2)^{\Nu-2}}
\sum_{j=1-\nu_1}^{\nu_2-1} (-1)^\Nu \, c^{(\Nu)}_{\nu_1,\nu_2;j}(d) \, \Big(\frac{m_1}{m_2}\Big)^j
\nonumber\\&+&
\frac{\intB^{1,0,1}_\mabc(d)}{(-m_1m_3)^{\Nu-2}}
\sum_{j=1-\nu_1}^{\nu_3-1} (-1)^\Nu \, c^{(\Nu)}_{\nu_1,\nu_3;j}(d) \, \Big(\!\!-\!\frac{m_1}{m_3}\Big)^j
\nonumber\\&+&
\frac{\intB^{0,1,1}_\mabc(d)}{(-m_2m_3)^{\Nu-2}}
\sum_{j=1-\nu_2}^{\nu_3-1} (-1)^\Nu \, c^{(\Nu)}_{\nu_2,\nu_3;j}(d) \, \Big(\!\!-\!\frac{m_2}{m_3}\Big)^j 
\ea
or equivalently, using \eq\nr{eq:masters} to make all mass dependence explicit,
\ba \la{eq:newconj2}
\frac{\intB^\nabc_\mabc(d)}{L^{2d}\,\Gamma^2(1-\frac{d}2)} &=& 
\sum_{j=1-\nu_1}^{\nu_2-1} (-1)^\Nu \, c^{(\Nu)}_{\nu_1,\nu_2;j}(d) \, m_1^{d-\Nu+j} \, m_2^{d-\Nu-j}
\nonumber\\&+&
\sum_{j=1-\nu_1}^{\nu_3-1} (-1)^j \, c^{(\Nu)}_{\nu_1,\nu_3;j}(d) \, m_1^{d-\Nu+j} \, m_3^{d-\Nu-j}
\nonumber\\&+&
\sum_{j=1-\nu_2}^{\nu_3-1} (-1)^j \, c^{(\Nu)}_{\nu_2,\nu_3;j}(d) \, m_2^{d-\Nu+j} \, m_3^{d-\Nu-j}
\;.
\ea
The coefficients $c^{(\Nu)}_{\nu_a,\nu_b;j}(d)$ are rational functions in $d$ that, in order to make the relation 
$\intB^\nabc_\mabc(d)=\intB^{\nu_2,\nu_1,\nu_3}_{m_2,m_1,m_3}(d)$ explicit, 
obey the symmetries (the second one being merely a special case of the first)
\ba \la{eq:csy}
c^{(\Nu)}_{\nu_a,\nu_b;j}(d) &=& c^{(\Nu)}_{\nu_b,\nu_a;-j}(d) \;,\quad
c^{(\Nu)}_{\nu_a,\nu_b;0}(d) = c^{(\Nu)}_{\nu_b,\nu_a;0}(d) \;.
\ea
The rational functions $c^{(\Nu)}(d)$ can e.g.\ be read off from IBP-generated reductions such as those given above (but see also our explicit solution given in \eq\nr{eq:kallen3} below).

\begin{figure}
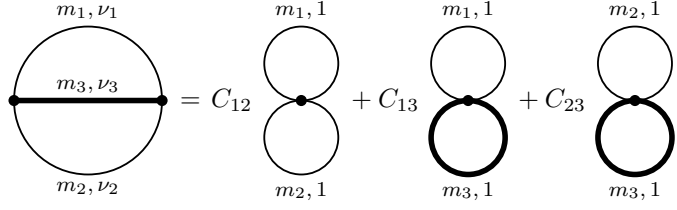

\setlength{\unitlength}{1.4pt}
\ba \nonumber
\sunB{$m_1,\nu_1$}{$m_2,\nu_2$}{$m_3,\nu_3$} &=&
C_{12}\,\tadA{$m_1,1$}{$m_2,1$}
+C_{13}\,\tadB{$m_1,1$}{$m_3,1$}
+C_{23}\,\tadB{$m_2,1$}{$m_3,1$}
\ea
\caption{\la{fig:conj}Structure of the general factorization formula \eq\nr{eq:newconj}. 
The $C_{ij}$ denote rational functions in masses $m_i, m_j$ and space-time dimension $d$ 
that depend on the (integer) values of the propagator powers $\nu_k$ that define the two-loop integral on the left-hand side.
For a special case, see \fig\ref{fig:B111}.}
\end{figure}

Equation \nr{eq:newconj} turns out to indeed be a generic representation of the reduced integral, as we will prove in the following \se\ref{se:conjProof}, starting from the IBP relation \eq\nr{eq:T92}.
For use in that proof, let us spell out the conjecture in the form of \eq\nr{eq:newconj2} for the special-mass case of \se\ref{se:B0mm}, where 
coefficient functions $\bzmm\nabc(d)$ had been introduced in \eqs\nr{eq:B0mmRes}-\nr{eq:B0mm}:
\ba \la{eq:new0mm}
\frac{\tB^\nabc(d)}{L^{2d}\,\Gamma^2(1-\frac{d}2)} &=&  \frac{\bzmm\nabc(d)}{m^{2\Nu-2d}}
\quad\Rightarrow\quad 
\sum_{j=1-\nu_2}^{\nu_3-1} (-1)^j\,c^{(\Nu)}_{\nu_2,\nu_3;j}(d) \;=\; \bzmm\nabc(d) \;.
\ea

We could use \eq\nr{eq:new0mm} together with the known result for $\tB^\nabc(d)$ to express the coefficients $c^{(\Nu)}(d)$ at $j=0$ in terms of the $j>0$ ones (keeping in mind \eq\nr{eq:csy}) as
\ba \la{eq:c0}
c^{(\Nu)}_{\nu_a,\nu_b;0}(d) &=& \bzmm{\Nu-\nu_a-\nu_b,\nu_a,\nu_b}
-\sum_{j=1}^{\nu_b-1} (-1)^j\,c^{(\Nu)}_{\nu_a,\nu_b;j}(d) 
-\sum_{j=1}^{\nu_a-1} (-1)^j\,c^{(\Nu)}_{\nu_b,\nu_a;j}(d)  \;,
\ea
where $\bzmm{}$ had been defined in \eq\nr{eq:B0mm}.

Expanding the binomial in the first line of \eq\nr{eq:BacaM} as $\frac{(m_3-m_1)^{2k}}{(-m_1m_3)^k}=\sum_{j=-k}^k\bin{2k}{j+k}\big(-\frac{m_1}{m_3}\big)^{j}$, shifting $j\rightarrow k+j$ and $j\rightarrow k-j$ in the two terms of the second line, respectively, and exchanging summations, one can compare the coefficients of the three master integrals with those of \eq\nr{eq:newconj} and read off
\ba\la{eq:caab1}
c^{(2\ka+\nu)}_{\ka,\ka;j}(d) &=&
\sum_{k=|j|}^{\ka-1} \bin{2k}{j+k} \frac{(-1)^{2\ka+\nu-k}\po{\nu}{k} \po{1-\tfrac{d}2}{\ka+\nu-1} \po{1+\nu+k-\tfrac{d}2}{\ka-1}}
{(\ka-1)!\,k!\,(\ka-1-k)!\,\po{2\ka+\nu+k-d}{\nu+k}} 
\;,
\\
c^{(2\ka+\nu)}_{\ka,\nu;j}(d) &=&
\sum_{k={\rm max}(0,-j)}^{\ka-1}
\frac{(-1)^{1+k+j} \po{\nu}{k} \po{1-\tfrac{d}2}{\ka+\nu-1} \po{1+\nu+k-\tfrac{d}2}{\ka-1}}{(k+j)!\,(\ka-1)!\,k!\,(\ka-1-k)!\,\po{2\ka+\nu+2k+j-d}{\nu-j}}
\;,
\\\la{eq:caab3}
c^{(2\ka+\nu)}_{\nu,\ka;j}(d) &=&
\sum_{k={\rm max}(0,j)}^{\ka-1}
\frac{(-1)^{1+k-j} \po{\nu}{k} \po{1-\tfrac{d}2}{\ka+\nu-1} \po{1+\nu+k-\tfrac{d}2}{\ka-1}}{(k-j)!\,(\ka-1)!\,k!\,(\ka-1-k)!\,\po{2\ka+\nu+2k-j-d}{\nu+j}}
\;.
\ea
As a check, note that the last two equations obey the symmetry relation \eq\nr{eq:csy}, as they should.
As another check, at $\ka=\nu$ all three expressions (while looking wildly different) give the same results for $c^{(3\nu)}_{\nu,\nu;j}(d)$.

%
\subsection{Proof of conjecture \eq\nr{eq:newconj}}
\la{se:conjProof}

Our proof of the conjecture \eq\nr{eq:newconj} will proceed via induction over the weight $\Nu=\nu_1+\nu_2+\nu_3$. 
From \eq\nr{eq:111} (or \nr{eq:B111}) it is clear that the conjecture holds for weight $\Nu=3$ (and we can read off $c^{(3)}_{1,1;0}(d)=-\frac{d-2}{2(d-3)}$).
Now suppose the conjecture holds at weight $(\Nu-1)$. We will show that this implies that it then also holds at weight $\Nu$ 
(and we can read off the $c^{(\Nu)}_{\nu_a,\nu_b;j}(d)$).

Starting from the IBP recurrence relation \eq\nr{eq:T92} (in this section only, we abbreviate shifted indices as $\bar1\equiv\nu_1-1$ etc., 
and use the two combinations $d_1\equiv d+2-\Nu$ and $d_2\equiv d+3-2\Nu$ to render expressions more compact)
\ba
\intB^\nabc_\mabc(d) \;\equiv\;
\intBi^{123} &=& -\frac1{2d_2m_1m_2m_3} \Big\{
[1]\intBi^{\bar123}
+[2]\intBi^{1\bar23}
+[3]\intBi^{12\bar3} 
\Big\} 
\ea
with
$[1]\equiv[m_1d_1+m_2\nu_3-m_3\nu_2],
~[2]\equiv[m_1\nu_3+m_2d_1-m_3\nu_1],
~[3]\equiv[m_1\nu_2+m_2\nu_1-m_3d_1]$,
we are allowed to use the conjecture \eq\nr{eq:newconj} on the rhs, since it contains integrals of weight $(\Nu-1)$ only. Extending notation by defining 
\ba
\la{eq:ext}
c^{(\Nu)}_{\nu_a,\nu_b;\nu_b} &\equiv& 0
\ea
in order to unify summation limits and re-arranging terms, this results in
\ba \la{eq:recu+conj}
\frac{2d_2\,\intBi^{123}}{(-1)^{\Nu}} &=& 
\frac{\intB^{1,1,0}(d)}{(m_1m_2)^{\Nu-2}}
\sum_{j=1-\nu_1}^{\nu_2-1} 
\Big(\frac{m_1}{m_2}\Big)^j \frac{(-1)}{(-m_3)} 
\Big[
[1]c^{(\Nu-1)}_{\bar1,2;j}
+[2]c^{(\Nu-1)}_{1,\bar2;j}
+[3]c^{(\Nu-1)}_{1,2;j}
\Big]
\nonumber\\&+&
\frac{\intB^{1,0,1}(d)}{(-m_1m_3)^{\Nu-2}}
\sum_{j=1-\nu_1}^{\nu_3-1} 
\Big(-\frac{m_1}{m_3}\Big)^j \frac{(-1)}{m_2} 
\Big[
[1]c^{(\Nu-1)}_{\bar1,3;j}
+[2]c^{(\Nu-1)}_{1,3;j}
+[3]c^{(\Nu-1)}_{1,\bar3;j}
\Big]
\nonumber\\&+&
\frac{\intB^{0,1,1}(d)}{(-m_2m_3)^{\Nu-2}}
\sum_{j=1-\nu_2}^{\nu_3-1} 
\Big(-\frac{m_2}{m_3}\Big)^j \frac{(-1)}{m_1} 
\Big[
[1]c^{(\Nu-1)}_{2,3;j}
+[2]c^{(\Nu-1)}_{\bar2,3;j}
+[3]c^{(\Nu-1)}_{2,\bar3;j}
\Big]
\;.\quad
\ea

In order to show that the conjecture holds at weight $\Nu$, we need to demonstrate that the 'wrong mass' in the denominator in front of each large square bracket cancels out. To this end, consider e.g.\ the first line of \eq\nr{eq:recu+conj}, where we need to factor a term $m_3$ out of the large square bracket; 
each small square bracket is linear in mass; we rewrite the three masses therein in terms of only $m_3$ and $(m_1-m_2)$, collect terms and obtain
\ba
\Big[\dots\Big]_{\stackrel{\mbox{\tiny\nr{eq:recu+conj}}}{\rm line1}} &=& \Big[
\frac{m_3}2\,y_{\nu_1,\nu_2,\nu_3;j}
+\frac{m_1-m_2}2\,(-1)^j\,z_{\nu_1,\nu_2,\nu_3;j}
\Big] \;,\\
y_{\nu_1,\nu_2,\nu_3;j} &\equiv& \big\{
(d_1-2\nu_2+\nu_3)c^{(\Nu-1)}_{\bar1,2;j}
+(d_1-2\nu_1+\nu_3)c^{(\Nu-1)}_{1,\bar2;j}
+(\Nu-\nu_3-2d_1)c^{(\Nu-1)}_{1,2;j}
\big\} \;,\qquad\\ 
\la{eq:defz}
z_{\nu_1,\nu_2,\nu_3;j} &\equiv& (-1)^j\,\big\{
(d_1-\nu_3)\big[c^{(\Nu-1)}_{\bar1,2;j}-c^{(\Nu-1)}_{1,\bar2;j}\big]
+(\nu_2-\nu_1)c^{(\Nu-1)}_{1,2;j}
\big\} \;,
\ea
where $y$ and $z$ are rational functions in $d$ (and hence do not contain masses).
We repeat the same exercise for the second (this time re-writing masses in terms of $m_2$ and $m_3+m_1$) and third (using $m_1$ and $m_3+m_2$) lines of \eq\nr{eq:recu+conj}, which produces a very similar structure
\ba
\Big[\dots\Big]_{\stackrel{\mbox{\tiny\nr{eq:recu+conj}}}{\rm line2}} &=& \Big[
-\frac{m_2}2\,y_{\nu_1,\nu_3,\nu_2;j}
+\frac{m_3+m_1}2\,(-1)^j\,z_{\nu_1,\nu_3,\nu_2;j}
\Big] \;,\\
\Big[\dots\Big]_{\stackrel{\mbox{\tiny\nr{eq:recu+conj}}}{\rm line3}} &=& \Big[
-\frac{m_1}2\,y_{\nu_2,\nu_3,\nu_1;j}
+\frac{m_3+m_2}2\,(-1)^j\,z_{\nu_2,\nu_3,\nu_1;j}
\Big] \;.
\ea
The integral hence naturally splits into two parts (we again abbreviate some indices writing 
$y_{\nu_1,\nu_2,\nu_3;j}=y_{1,2,3;j}$ henceforth, same for the functions $z$),
\ba
\intBi^{123} &=& 
\intBi^{123}_y
+\intBi^{123}_z \\
\la{eq:B123y}
\frac{4d_2\,\intBi^{123}_y}{(-1)^{\Nu} } &=& 
\frac{\intB^{1,1,0}(d)}{(m_1m_2)^{\Nu-2}}
\sum_{j=1-\nu_1}^{\nu_2-1} 
\Big(\frac{m_1}{m_2}\Big)^j y_{1,2,3;j}
\nonumber\\&+&
\frac{\intB^{1,0,1}(d)}{(-m_1m_3)^{\Nu-2}}
\sum_{j=1-\nu_1}^{\nu_3-1} 
\Big(-\frac{m_1}{m_3}\Big)^j y_{1,3,2;j}
\nonumber\\&+&
\frac{\intB^{0,1,1}(d)}{(-m_2m_3)^{\Nu-2}}
\sum_{j=1-\nu_2}^{\nu_3-1} 
\Big(-\frac{m_2}{m_3}\Big)^j y_{2,3,1;j}
\\
\la{eq:Bz}
\frac{4d_2\,\intBi^{123}_z}{(-1)^{\Nu}} &=& 
\frac{\intB^{1,1,0}(d)}{(m_1m_2)^{\Nu-2}}
\frac{m_2-m_1}{(-m_3)}
\sum_{j=1-\nu_1}^{\nu_2-1} 
\Big(-\frac{m_1}{m_2}\Big)^j 
z_{1,2,3;j}
\nonumber\\&+&
\frac{\intB^{1,0,1}(d)}{(-m_1m_3)^{\Nu-2}}
\frac{m_3+m_1}{(-m_2)}
\sum_{j=1-\nu_1}^{\nu_3-1} 
\Big(\frac{m_1}{m_3}\Big)^j 
z_{1,3,2;j}
\nonumber\\&+&
\frac{\intB^{0,1,1}(d)}{(-m_2m_3)^{\Nu-2}}
 \frac{m_3+m_2}{(-m_1)}
\sum_{j=1-\nu_2}^{\nu_3-1} 
\Big(\frac{m_2}{m_3}\Big)^j 
z_{2,3,1;j}
\ea
While we are happy with $\intBi^{123}_y$ (as it already has the form of the weight-($\Nu$) conjecture), the $z$\/-terms need further work. 
The idea is to subtract a suitably constructed zero. The key observation is 
\ba
\sum_{j=1-a}^{b-1} z_{a,b,c;j} 
&=&
(d_1-c)\big[\beta^{c,\bar{a},b}-\beta^{c,a,\bar{b}}\big]
+(b-a) \beta^{\bar{c},a,b}  \\
&=&
\frac{2\,(d_1-c)\,(d_1-c-1)\,\beta^{c,a,b}}{(d_1-1)(d+2-2b-2c)(d+2-2a-2c)}
\times\\&&\times\nonumber
\la{eq:sqb}
\Big[(1-a)(d+2-2b-2c)-(1-b)(d+2-2a-2c)-(b-a)(d-2c)\Big]
\\&=&0 \;,
\ea
where we have first used \eqs\nr{eq:defz} and \nr{eq:new0mm} and then employed the identities
\ba
\la{eq:beta1bar}
\frac{\beta^{\bar1,2,3}}{\beta^{1,2,3}} &=& \frac{-2(d-2\nu_1)(d+1-\Nu-\nu_1)(d+2-\Nu-\nu_1)}{(d+2-2\nu_1-2\nu_2)(d+2-2\nu_1-2\nu_3)(d+1-\Nu)} \mbox{~~is the recursion \eq\nr{eq:T93}} \\
\frac{\beta^{1,\bar2,3}}{\beta^{1,2,3}} &=& \frac{2(1-\nu_2)(d+1-\Nu-\nu_1)}{(d+2-2\nu_1-2\nu_2)(d+1-\Nu)} \mbox{~~follows from $\Gamma$'s, see \eq\nr{eq:B0mm}} \\
\la{eq:beta3bar}
\frac{\beta^{1,2,\bar3}}{\beta^{1,2,3}} &=& \frac{2(1-\nu_3)(d+1-\Nu-\nu_1)}{(d+2-2\nu_1-2\nu_3)(d+1-\Nu)} 
\mbox{~~via $2\leftrightarrow3$ symmetry of $\beta^{1,2,3}$}
\ea
to shift the $\beta$ to a common set of indices, after which the square bracket of \eq\nr{eq:sqb} is seen to be identically zero. 
This allows us to subtract three vanishing terms  from \eq\nr{eq:Bz}, by replacing $(X)^j\rightarrow(X)^j-1$ under each $j$\/-sum. Now, the $j=0$ term of each sum does not contribute, and we 
replace $j\rightarrow-j$ for negative summation indices. This results in the intermediate expression
\ba
\frac{4d_2\,\intBi^{123}_z}{(-1)^{\Nu}} &=& 
\frac{\intB^{1,1,0}(d)}{(m_1m_2)^{\Nu-2}}
\frac{m_2-m_1}{(-m_3)}
\Big\{
\sum_{j=1}^{\nu_2-1} 
\Big[
\Big(-\frac{m_1}{m_2}\Big)^j 
-1 \Big]
z_{1,2,3;j}
+\sum_{j=1}^{\nu_1-1} 
\Big[
\Big(-\frac{m_2}{m_1}\Big)^j 
-1 \Big]
z_{1,2,3;-j}
\Big\}
\nonumber\\&+&
\frac{\intB^{1,0,1}(d)}{(-m_1m_3)^{\Nu-2}}
\frac{m_3+m_1}{(-m_2)}
\Big\{
\sum_{j=1}^{\nu_3-1} 
\Big[
\Big(\frac{m_1}{m_3}\Big)^j 
-1 \Big]
z_{1,3,2;j}
+\sum_{j=1}^{\nu_1-1} 
\Big[
\Big(\frac{m_3}{m_1}\Big)^j 
-1 \Big]
z_{1,3,2;-j}
\Big\}
\nonumber\\&+&
\frac{\intB^{0,1,1}(d)}{(-m_2m_3)^{\Nu-2}}
\frac{m_3+m_2}{(-m_1)}
\Big\{
\sum_{j=1}^{\nu_3-1} 
\Big[
\Big(\frac{m_2}{m_3}\Big)^j 
-1 \Big]
z_{2,3,1;j}
+\sum_{j=1}^{\nu_2-1} 
\Big[
\Big(\frac{m_3}{m_2}\Big)^j 
-1 \Big]
z_{2,3,1;-j}
\Big\}
\;.\nonumber
\ea
For each difference in square brackets we can now use the truncated geometric series 
\ba
\big[x^j-1\big] &=& (x-1) \sum_{k=0}^{j-1} x^k
\ea
in order to factor off one term (note that this step requires $j>0$, justifying the split performed above).
For example, 
$[(\frac{m_3}{m_2})^j-1]=(\frac{m_3}{m_2}-1)\sum_{k=0}^{j-1}(\frac{m_3}{m_2})^k$
for the last term, after which $(\frac{m_3}{m_2}-1)=\frac{m_1}{m_2}$
makes the desired cancellation against the prefactor mass $\frac1{m_1}$ explicit.
This cancellation indeed happens for all six terms, such that we now have
\ba
\frac{4d_2\,\intBi^{123}_z}{(-1)^{\Nu}} &=& 
\frac{\intB^{1,1,0}(d)}{(m_1m_2)^{\Nu-2}}
\Big\{
\sum_{j=1}^{\nu_2-1} 
\sum_{k=0}^{j-1}
\Big(-\frac{m_1}{m_2}\Big)^k
\big(1-\tfrac{m_1}{m_2}\big)
z_{1,2,3;j}
-\sum_{j=1}^{\nu_1-1}  
\sum_{k=0}^{j-1}
\Big(-\frac{m_2}{m_1}\Big)^k
\big(1-\tfrac{m_2}{m_1}\big)
z_{1,2,3;-j}
\Big\}
\nonumber\\&+&
\frac{\intB^{1,0,1}(d)}{(-m_1m_3)^{\Nu-2}}
\Big\{
\sum_{j=1}^{\nu_3-1}  
\sum_{k=0}^{j-1}
\Big(\frac{m_1}{m_3}\Big)^k
\big(1+\tfrac{m_1}{m_3}\big)
z_{1,3,2;j}
-\sum_{j=1}^{\nu_1-1}  
\sum_{k=0}^{j-1}
\Big(\frac{m_3}{m_1}\Big)^k
\big(1+\tfrac{m_3}{m_1}\big)
z_{1,3,2;-j}
\Big\}
\nonumber\\&+&
\frac{\intB^{0,1,1}(d)}{(-m_2m_3)^{\Nu-2}}
\Big\{
\sum_{j=1}^{\nu_3-1}  
\sum_{k=0}^{j-1}
\Big(\frac{m_2}{m_3}\Big)^k
\big(1+\tfrac{m_2}{m_3}\big)
z_{2,3,1;j}
-\sum_{j=1}^{\nu_2-1}  
\sum_{k=0}^{j-1}
\Big(\frac{m_3}{m_2}\Big)^k 
\big(1+\tfrac{m_3}{m_2}\big)
z_{2,3,1;-j}
\Big\}
\;,\nonumber
\ea
which already has the form of the conjecture and completes the proof. To be completely explicit, let us absorb the extra mass factors into the $k$\/-sum and interchange summations
\ba
\sum_{k=0}^{j-1}x^k(1+x) &=& \sum_{k=0}^{j}x^k(2-\delta_k-\delta_{k-j}) \;,\quad
\sum_{j=1}^{\nu_i-1} \sum_{k=0}^{j} =
\sum_{k=0}^{\nu_i-1} \sum_{j={\rm max}(k,1)}^{\nu_i-1} \;,
\ea
followed by re-labelling summation indices $j\leftrightarrow k$, to get, 
using $\djk\equiv(2-\delta_j-\delta_{j-k})$ and $j_1\equiv{\rm max}(j,1)$,
\ba \la{eq:B123z}
\frac{4d_2\,\intBi^{123}_z}{(-1)^{\Nu}} &=& 
\frac{\intB^{1,1,0}(d)}{(m_1m_2)^{\Nu-2}}
\Big\{
\sum_{j=0}^{\nu_2-1} 
\sum_{k=j_1}^{\nu_2-1}
\Big(-\frac{m_1}{m_2}\Big)^j
\djk\,
z_{1,2,3;k}
-\sum_{j=0}^{\nu_1-1}  
\sum_{k=j_1}^{\nu_1-1}
\Big(-\frac{m_2}{m_1}\Big)^j
\djk\,
z_{1,2,3;-k}
\Big\}
\nonumber\\&+&
\frac{\intB^{1,0,1}(d)}{(-m_1m_3)^{\Nu-2}}
\Big\{
\sum_{j=0}^{\nu_3-1}  
\sum_{k=j_1}^{\nu_3-1}
\Big(\frac{m_1}{m_3}\Big)^j
\djk\,
z_{1,3,2;k}
-\sum_{j=0}^{\nu_1-1}  
\sum_{k=j_1}^{\nu_1-1}
\Big(\frac{m_3}{m_1}\Big)^j
\djk\,
z_{1,3,2;-k}
\Big\}
\nonumber\\&+&
\frac{\intB^{0,1,1}(d)}{(-m_2m_3)^{\Nu-2}}
\Big\{
\sum_{j=0}^{\nu_3-1}  
\sum_{k=j_1}^{\nu_3-1}
\Big(\frac{m_2}{m_3}\Big)^j
\djk\,
z_{2,3,1;k}
-\sum_{j=0}^{\nu_2-1}  
\sum_{k=j_1}^{\nu_2-1}
\Big(\frac{m_3}{m_2}\Big)^j
\djk\,
z_{2,3,1;-k}
\Big\}
\;.
\ea

%
\section{Two-step IBP recursion relation}
\la{se:twoStep}

Motivated by the fact that in \se\ref{se:Baac} the relation \eq\nr{eq:T94b} involving dimensional shifts had been instrumental 
in generalizing $\intB^{1,1,\nu}(d)$ to $\intB^{\ka,\ka,\nu}(d)$, let us take a closer look at further such dimension-shift relations. 
In our notation, the first and third of \eqs(94) of (the journal version of) \cite{Tarasov:1997kx} read 
\ba
\la{eq:T94a} (d-2)(1-\nu_1)\intB^\nabc(d) &=& \Big[2m_2m_3+\subA-\subB-\subC\Big]\,\subA\,\subDD\,\intB^\nabc(d)\;,\\
\la{eq:T94c} (d-2)(d-\Nu) \intB^\nabc(d) &=& \Big[-2m_2m_3\subA-2m_3m_1\subB+2m_1m_2\subC\Big]\,\subDD\,\intB^\nabc(d)\;,
\ea
where $\subDD$ is the dimension-shifting operator of \eq\nr{eq:subDD}.
An interesting combination of these two IBP relations is obtained by adding \eq\nr{eq:T94a} plus its two 
permutations\footnote{To not miss signs when permuting indices, it is useful to remember that $2m_2m_3=-m_1^2+m_2^2+m_3^2$.} 
to \eq\nr{eq:T94c}, whereupon the mass-dependent terms drop out, leaving
\ba \la{eq:AD82}
(d-2)(d+3-2\Nu)\intB^\nabc(d) &=& \lambda(\subA,\subB,\subC)\,\subDD\,\intB^\nabc(d)\\
\mbox{with~~~} \lambda(\subA,\subB,\subC) &=& \subA\subA + \subB\subB + \subC\subC -2(\subA\subB+\subB\subC+\subC\subA) \;,
\ea
where we once again encounter K\"all\'en's triangle function of \eq\nr{eq:defKallen}, this time in operator form, 
as well as the dimension-lowering operator $\subDD$ of \eq\nr{eq:subDD}.

Noting that \eq\nr{eq:AD82} reduces the index-weight $\Nu=\nu_1+\nu_2+\nu_3$ by two in each step, and given that we know 
the boundary integrals \eqs\nr{eq:B120}-\nr{eq:bc4} for arbitrary dimension $d$, it might be used in place of the single-step 
recursion \eq\nr{eq:T92} to construct a more efficient recursive reduction to master integrals. 
In practice, starting from an integral $\intB^\nabc(d)$ with positive indices $\nu_i$, one can repeatedly apply \eq\nr{eq:AD82} 
until one of the indices is reduced to 0 or -1. The latter case is then reduced to the first by using the denominator's symmetry 
(under momentum shifts $p\rightarrow-p$, $q\rightarrow2p-q$ and $p\rightarrow2q-p$ for $\intB^{a,b,0}$, $\intB^{a,0,c}$ and $\intB^{0,b,c}$, 
respectively) in the integral's numerator (see also \eqs(2.9), (2.10) of \cite{ Davydychev:1992mt})
\ba
\la{eq:subC}\subC\,\intB^{\nu_1,\nu_2,0}(d) &=& \Big\{2m_1m_2+\subA+\subB\Big\}\,\intB^{\nu_1,\nu_2,0}(d) \;,\\
\la{eq:subB}\subB\,\intB^{\nu_1,0,\nu_3}(d) &=& \Big\{-2m_1m_3+\subA+\subC\Big\}\,\intB^{\nu_1,0,\nu_3}(d) \;,\\
\la{eq:subA}\subA\,\intB^{0,\nu_2,\nu_3}(d) &=& \Big\{-2m_2m_3+\subB+\subC\Big\}\,\intB^{0,\nu_2,\nu_3}(d) \;,
\ea
where we have exploited $m_3=m_1+m_2$ in order to write differences of squared masses in a convenient way.
Now one is left with products of tadpoles, for which \eqs\nr{eq:bc1}-\nr{eq:bc4} provide the reduction to master integrals.

%
\subsection{Recursion-free result for $c^{(\Nu)}$}
\la{se:4.1}

It turns out that the two-step reduction \eq\nr{eq:AD82} together with \eqs\nr{eq:subC}-\nr{eq:subA} 
not only allows for a very efficient recursive calculation of integrals $\intB^\nabc(d)$ with high values of the indices $\nu_i$,
but also allows to explicitly construct the rational functions $c^{(\Nu)}(d)$, as we will explain in this section.

Let us first simplify the recurrence relation \eq\nr{eq:AD82}, to clearly expose its structure. 
We start by observing that the recurrence leaves the combination $d-\Nu\equiv \dc$ invariant, 
and use this to replace $d=\dc+\Nu$, substituting back only {\em after} solving the equation. 
Furthermore, let us define $e\equiv{\rm mod}(\Nu,2)$ (i.e.\ $e=0/1$ for even/odd index sum $\Nu$; 
$e$ is not changed by the recursion either, since it lowers the index sum by two at each step), 
as well as the integer $\Nn\equiv\ceil{\frac{\Nu}2}=\frac{\Nu+e}2$, which we use to set $\Nu=2\Nn-e$. 
With these definitions, \eq\nr{eq:AD82} can be written as (recall $L^4=(4\pi)^{-2}$)
\ba
a_\Nn\,\intB^\nabc_\Nn &=& \lambda(\subA,\subB,\subC)\,L^4\,\intB^\nabc_{\Nn-1} 
\quad\mbox{with}\quad a_\Nn=(\dc-e-2+2\Nn)(\dc+e+3-2\Nn) \;,\qquad
\ea
where the lowering operator arguments of the \kallen polynomial $\lambda(\subA,\subB,\subC)$ act upon the indices $\nu_i$ 
on the right-hand side, while $\Nu=\nu_1+\nu_2+\nu_3$ and therefore the value of $\Nn$ is fixed on both sides.
To absorb the dimension-lowering operator that comes with each application of $\lambda$ in \eq\nr{eq:AD82}, 
we have slightly modified the notation for the integrals $\intB$ and absorbed the dimension-shift into the index $n_0$.

To remove the factor $a_\Nn$ from the left-hand side, we can define (recalling \eq\nr{eq:defPoch})
\ba\la{eq:bintBdef}
\bintB^\nabc_\Nn &\equiv& \bigg(\prod_{j=1}^\Nn \frac{a_j}{L^4}\bigg)\,\intB^\nabc_\Nn \;=\; 
\big(-\tfrac{4}{L^4}\big)^\Nn\,\po{\tfrac{\dc-e}2}{\Nn}\,\po{-\tfrac{\dc+e+1}2}{\Nn}\,\intB^\nabc_\Nn \;,
\ea
in terms of which we obtain a 3-parameter linear homogeneous recurrence with constant coefficients,
\ba\la{eq:cleanRecu}
\bintB^\nabc_\Nn &=& \lambda(\subA,\subB,\subC)\,\bintB^\nabc_{\Nn-1} \;.
\ea

Starting from an integral $\bintB^\nabc$ with all $\nu_i$ positive, one can now use \eq\nr{eq:cleanRecu} until one of the $\nu_i$ is 
either 0 or $-1$. It can happen that two indices get reduced to zero simultaneously, in which case the boundary 
condition \eq\nr{eq:bc1} allows us to drop these integrals. The recursion hence terminates with a linear combination of integrals in 
which exactly one index is $0$ or $-1$, while the two others remain positive, dividing the result into three sectors. 
Due to \eq\nr{eq:newconj}, to fix the coefficient functions $c^{(\Nu)}(d)$ therein it is sufficient to focus on only one of those sectors here. 
We choose this to be the one with non-positive $\nu_3$, corresponding to tadpole integrals in the masses $m_1$ and $m_2$, or the first lines of \eqs\nr{eq:newconj} and \nr{eq:newconj2}. In this sector, the recursion terminates with
\ba\la{eq:Bsum}
\bintB^\nabc &=& \sum_{j=1}^{\nu_1}\sum_{k=1}^{\nu_2} \cc^\nabc_{j,k;p}\,\bintB^{j,k,-p} + \mbox{other sectors} 
\;,\quad\mbox{with}\quad p\equiv{\rm mod}\big(\Nu-j-k,2\big) \;,\quad
\ea
where the $\cc^\nabc_{j,k;p}$ are some integer coefficients generated by the recursion.

To understand the integer coefficients $\cc^\nabc_{j,k;p}$, it helps to backtrack the recurrence one step. 
We need to distinguish the two terminal cases $p\in\{0,1\}$.
If $p=1$, i.e.\ $\bintB^{j,k,-1}$ was produced, it must have originated from the $[\subC]^2$ part of $\lambda$, as $[\subC]^2\,\bintB^{j,k,1}$. 
The positive-index integral $\bintB^{j,k,1}$, in turn, arose from $\lambda^{\frac{\Nu-j-k-1}2}\,\bintB^\nabc$ by picking out the 
term $[\subA]^{\nu_1-j}\,[\subB]^{\nu_2-k}\,[\subC]^{\nu_3-1}$ from the product of $\lambda$'s. Denoting the (integer) coefficient of that 
term as $g_{\nu_1-j,\nu_2-k,\nu_3-1}$, we have just learned that 
\ba
\cc^\nabc_{j,k;1}=g_{\nu_1-j,\nu_2-k,\nu_3-1} \;,
\ea
where $g_{a,b,c}$ is the coefficient of the term $x^ay^bz^c$ in the expanded product $\big[\lambda(x,y,z)\big]^{\frac{a+b+c}2}$ or, equivalently, in the series expansion of $\frac1{1-\lambda(x,y,z)}$, which can therefore be thought of as the generating function for the $g_{a,b,c}$
\ba\la{eq:gabcDef}
\sum_{a,b,c\ge0} g_{a,b,c}\,x^a y^b z^c = \frac1{1-\lambda(x,y,z)} = \frac1{1-x^2-y^2-z^2+2xy+2xz+2yz} \;.
\ea

If on the other hand $p=0$,  i.e.\ $\bintB^{j,k,0}$ was produced, it can have come from the part $[\subC]^2-2(\subA\subC+\subB\subC)$ of $\lambda$, 
as $[\subC]^2\,\bintB^{j,k,2}-2\,\subA\subC\,\bintB^{j+1,k,1}-2\,\subB\subC\,\bintB^{j,k+1,1}$ (the first term obviously only contributes 
for $\nu_3>1$). Each of these positive-index integrals $\bintB$, in turn, arose from the respective powers of lowering operators 
within $\lambda^{\frac{\Nu-j-k-2}2}\,\bintB^\nabc$, whose coefficients we have already named above as $g$. So we have now learned that 
\ba\la{eq:cnabc1}
\cc^\nabc_{j,k;0}&=&g_{\nu_1-j,\nu_2-k,\nu_3-2}-2\,g_{\nu_1-j-1,\nu_2-k,\nu_3-1}-2\,g_{\nu_1-j,\nu_2-k-1,\nu_3-1} \\
&=&\cc^{\nu_1,\nu_2,\nu_3-1}_{j,k;1}-2\,\cc^{\nu_1,\nu_2,\nu_3}_{j+1,k;1}-2\,\cc^{\nu_1,\nu_2,\nu_3}_{j,k+1;1}\;.
\ea

The integers $g_{a,b,c}$ as defined in \eq\nr{eq:gabcDef} are obviously fully symmetric in all three indices, and we will derive a closed form for them later (cf.\ \eq\nr{eq:gResult} below). In the above, these coefficients are needed for non-negative indices with even sum $a+b+c$. We have $g_{a,b,c}=0$ for all other cases (i.e.\ if any index is negative, or the index sum is odd). This automatically implements the upper summation limits of \eq\nr{eq:Bsum}, for example, and also guarantees that the first term of \eq\nr{eq:cnabc1} only contributes for $\nu_3>0$, such that we did not need to enforce this with an additional condition in the equation itself.

Translating back from integrals $\bintB$ to $\intB$ via \eq\nr{eq:bintBdef} and recalling the parameter definitions we had made above, we now have
\ba\la{eq:translated}
\intB^\nabc(d) &=& \sum_{j=1}^{\nu_1}\sum_{k=1}^{\nu_2} \cc^\nabc_{j,k;p}\,\frac{L^{4n}\,\intB^{j,k,-p}(d-2n)}{\qp^\Nu_n(d)} + \mbox{other sectors}\;,\\
\mbox{with} &&
p\equiv{\rm mod}(\Nu-j-k,2) \;,\quad
n\equiv\frac{\Nu-j-k+p}2 \\
\mbox{and}&&
\qp^\Nu_n(d)\equiv\Big(\prod_{\ell=1}^{n}(d-2\ell)(d+1-2\Nu+2\ell)\Big) \;.
\ea

For $p=1$, we remove the negative index from the integrals $\intB^{j,k,-1}(d-2n)$ via \eq\nr{eq:subC}, such that only integrals $\intB^{a,b,0}(d-2n)$ remain. Those are known analytically, cf.\  \eq\nr{eq:B120}, and can for example all be mapped onto one common master integral $\intB^{1,1,0}(d)$ as given in \eq\nr{eq:masters}, using
\ba\la{eq:map}
\intB^{a,b,0}(d-2n) &=& \frac{h_{a,n}(d)\,h_{b,n}(d)\,\intB^{1,1,0}(d)}{m_1^{2a+2n-2}\,m_2^{2b+2n-2}\,L^{4n}} \;,\quad\mbox{with}\quad 
h_{a,n}(d)\;\stackrel{\nr{eq:defh}}=\;\frac{\Gamma(a+n-\frac{d}2)}{\Gamma(a)\,\Gamma(1-\frac{d}2)} \;.
\ea
Equation \nr{eq:translated} (modulo the other sectors with $\nu_2=0$ and $\nu_1=0$, respectively) then becomes
\ba
\intB^\nabc_{{\rm (sector 1)}}(d) &=& \frac{\intB^{1,1,0}(d)}{(m_1m_2)^{\Nu-2}}
\sum_{j=1}^{\nu_1}\sum_{k=1}^{\nu_2} \frac{h_{j,n}(d)\,h_{k,n}(d)}{\qp^\Nu_n(d)\,(m_1m_2)^p}\,\Big(\frac{m_1}{m_2}\Big)^{k-j}
\Bigg\{\begin{array}{rl}
p=0: & \cc^\nabc_{j,k;0} \\
p=1: & \cc^\nabc_{j,k;1} \,\ct
\end{array}
\;,
\ea
where $\ct \equiv \big[2m_1m_2+m_1^2\,\frac{h_{j-1,n}(d)}{h_{j,n}(d)}+m_2^2\,\frac{h_{k-1,n}(d)}{h_{k,n}(d)}\big]$.
The fractions (like $\frac{h_{j-1,n}(d)}{h_{j,n}(d)}=\frac{2(1-j)}{(d+2-2j-2n)}$) arose from factoring $\intB^{j,k,0}(d)$ out of all terms.

We now trade the summation parameter $k$ for $\ell\equiv k-j$, in effect summing over diagonal stripes of the $(j,k)$\/-rectangle, leading to
\ba
\intB^\nabc_{{\rm (sector 1)}}(d) &=& \frac{\intB^{1,1,0}(d)}{(m_1m_2)^{\Nu-2}}\sum_{\ell=1-\nu_1}^{\nu_2-1}\Big(\frac{m_1}{m_2}\Big)^\ell
\sum_{j={\rm max}(1,1-\ell)}^{{\rm min}(\nu_1,\nu_2-\ell)} \frac{h_{j,n}(d)\,h_{j+\ell,n}(d)}{\qp^\Nu_n(d)}
\Bigg\{\begin{array}{rl}
p=0: & \cc^\nabc_{j,j+\ell;0} \\
p=1: & \cc^\nabc_{j,j+\ell;1}\, \ch
\end{array}
\;,\qquad
\ea
where now $\ch \equiv \big[2+\frac{m_1}{m_2}\,\frac{h_{j-1,n}(d)}{h_{j,n}(d)}+\frac{m_2}{m_1}\,\frac{h_{j+\ell-1,n}(d)}{h_{j+\ell,n}(d)}\big]$.
In terms of the new summation parameters, $p={\rm mod}(\Nu-\ell,2)$ and $n=\frac{\Nu-\ell+p}2-j$.

In order to facilitate a comparison with \eq\nr{eq:newconj}, we collect same powers of the mass ratio by shifting summation indices. To this end, for the $\frac{m_1}{m_2}$ term we let $\ell\rightarrow\ell-1$ and $j\rightarrow j+1$ (and hence $p\rightarrow1-p$, $n\rightarrow n-p$), while the $\frac{m_2}{m_1}$ term necessitates $\ell\rightarrow\ell+1$ (and hence $p\rightarrow1-p$, $n\rightarrow n+1-p$). 
Recalling that $h_{a,n}(d)$ vanishes for $j\le0$ (cf.\ \eq\nr{eq:defh}), we do not need to adjust the limits of the $j$\/-sum.
The resulting expression reads
\ba
\intB^\nabc_{{\rm (sector 1)}}(d) &=& \frac{\intB^{1,1,0}(d)}{(m_1m_2)^{\Nu-2}}\sum_{\ell=1-\nu_1}^{\nu_2-1}\Big(\frac{m_1}{m_2}\Big)^\ell
\sum_{j={\rm max}(1,1-\ell)}^{{\rm min}(\nu_1,\nu_2-\ell)} \frac{h_{j,n}(d)\,h_{j+\ell,n}(d)}{\qp^\Nu_n(d)}
\Bigg\{\begin{array}{rl}
p=0: & \cb^{\,\nabc}_{j,j+\ell}\\
p=1: & 2\,\cc^\nabc_{j,j+\ell;1}
\end{array}
\;,\qquad
\ea
with $\cb^{\,\nabc}_{j,j+\ell} \equiv \cc^\nabc_{j,j+\ell;0} +\cc^\nabc_{j+1,j+\ell;1}+\cc^\nabc_{j,j+\ell+1;1}$.
We note that for actual polynomial expansions, the representation
$h_{a,n}(d)=\frac1{\Gamma(a)\,2^{a+n-1}}\,\prod_{\ell=1}^{a+n-1}(2\ell-d)$ turns out to be useful;
also, we observe that one actually never needs to explicitly use the trivial boundary values 
$\intB^{\nu_1,0,0}(d)=0$ etc.\ of \eq\nr{eq:bc1} here, since they always drop out exactly (after lifting the negative index).

In summary, \kallen recursion, lifting and tadpole reduction leads to ($n_j\equiv\ceil{\frac{\Nu+j}2}$ for brevity)
\ba\la{eq:kallen}
(-1)^\Nu\,c^{(\Nu)}_{\nu_a,\nu_b;j}(d) &=& 
\sum_{k={\rm max}(1+j,1)}^{{\rm min}(\nu_a+j,\nu_b)} 
\frac{\po{1-\frac{d}2}{n_j-1}\,\po{1-\frac{d}2}{n_j-j-1}}{\po{1-\frac{d}2}{n_j-k}\,\po{\frac{d+3}2-\Nu}{n_j-k}}\,
\frac{G_{\nu_a+j-k,\nu_b-k,\nu_c-1}}{(-4)^{n_j-k}\,\Gamma(k)\,\Gamma(k-j)} \;,
\ea
where the integer coefficients $G$ are given by 
\ba\la{eq:G}
G_{a,b,c} &=& 
\Bigg\{\begin{array}{lcl}
(a+b+c) \mbox{~even}& : & 2\,g_{a,b,c} \;,\\
(a+b+c) \mbox{~odd}& : & g_{a,b,c-1}-g_{a-1,b,c}-g_{a,b-1,c} \;,
\end{array}
\ea
with $g_{a,b,c}$ defined in \eq\nr{eq:gabcDef}.

As a check, the symmetry relation $c^{(\Nu)}_{\nu_b,\nu_a;-j}(d)=c^{(\Nu)}_{\nu_a,\nu_b;j}(d)$ (cf.\ \eq\nr{eq:csy}) follows from \eq\nr{eq:kallen} by 
simply shifting $k\rightarrow k-j$ and using that $G_{a,b,c}=G_{b,a,c}$ is symmetric in its first two indices.
Furthermore, the representation \eq\nr{eq:kallen} could now be used to render the proof of the previous section constructive.

%
\subsection{Closed form for integer coefficients $g_{a,b,c}$}
\la{se:4.2}

We now turn to deriving a closed form for the numbers $g_{a,b,c}$ that had been defined in \eq\nr{eq:gabcDef}.
The numbers $g_{a_1,a_2,a_3}$ are actually the solution of a purely combinatorial problem, counting distinct weighted shortest paths between two points in a 3d cubic grid, originating from the \kallen recursion. 
To evaluate $g_{a_1,a_2,a_3}$, count\footnote{We are indebted to Felix Galland for discussions on the 2d version, which led to the 3d solution presented here.} all possible words made out of three different letters (each letter representing one step on the grid into the direction labelled by that letter), taking $a_i$ copies of the $i$-\/th letter, respectively, but weighing each word with a factor of $(-1)^m$ where $m$ counts the mixed-letter pairs in the 
word.\footnote{For example, we can obtain $g_{1,1,2}$ by first writing all $\frac{(1+1+2)!}{1!\,1!\,2!}=12$ words made from the letters ABCC as 
$\{$ABCC,ACBC,ACCB,CABC,CACB,CCAB,BACC,BCAC,BCCA,CBAC,CBCA,CCBA$\}$; then generating the weights by grouping letters of each word in pairs of two, replacing same-letter (mixed-letter) pairs with a factor of $1$ ($-1$) and multiplying, getting $\{-1,1,1,1,1,-1,-1,1,1,1,1,-1\}$; and finally summing these weights, which results in $g_{1,1,2}=4$.}
The result can be represented as a triple sum that follows directly from counting the letter combinations as explained above. 
Using $A\equiv\frac{a_1+a_2+a_3}2$ and $N\equiv A-n_1-n_2-n_3$, one obtains 
\ba\la{eq:gTriple}
g_{a_1,a_2,a_3} &=& 
\sum_{n_1=0}^{\floor{\frac{a_1}2}}
\sum_{n_2=0}^{\floor{\frac{a_2}2}}
\sum_{n_3=0}^{\floor{\frac{a_3}2}}
\frac{(-2)^N\,A!}{n_1!\,n_2!\,n_3!\,(N+2n_1-a_1)!\,(N+2n_2-a_2)!\,(N+2n_3-a_3)!} \;.
\ea
This integer-valued triple sum $g_{a_1,a_2,a_3}$ is totally symmetric in its three indices. It is implied that the sum of indices, $a_1+a_2+a_3$, is always even, cf.\ \eq\nr{eq:gabcDef}.
Special cases are 
\ba\la{eq:gZero}
g_{a_1<0,a_2,a_3} = 0 \;,\quad
g_{0,0,a_3} = 1 \;,\quad
g_{0,a_2,a_3} = (-1)^{a_2}\frac{(a_2+a_3)!}{a_2!\,a_3!} \;,
\ea
the last two of which correspond to the respective 1d and 2d counting problems.

It would be great to have a closed form for the general case $g_{a,b,c}$, when all three indices are positive.
Not being able to explicitly solve \eqs\nr{eq:gabcDef} or \nr{eq:gTriple} for now, we can take another approach and compare our result 
for $\intB^\nabc(d)$ in terms of the $g_{a,b,c}$ of the previous section to special cases of $\intB$ that we had already been able to solve in closed form. As we will explain below, it turns out that these constraints are strong enough to extract an explicit formula for the $g_{a,b,c}$, which can then of course be verified against \eqs\nr{eq:gabcDef} or \nr{eq:gTriple}. A key realization is that, even though one is equating two sums of rational functions (in $d$), the pole structure of both expressions has to be identical, allowing to access single terms of those sums.

Let us\footnote{See also \app\ref{se:gAlt}, however.} compare \eq\nr{eq:kallen} for the special case $\nu_a=\nu_b\equiv\ka$ with our solution for 
$\intB^{\ka,\ka,\nu_c}(d)$ (from which we had read off $c^{(2\ka+\nu_c)}_{\ka,\ka;j}(d)$ as a finite sum over Gamma functions, cf.\ \eq\nr{eq:caab1}), 
on the single pole $\frac1{d-p}$ with $p=2(2\ka+\nu_c+k-n_j)-1$ where $k={\rm max}(1+j,1)$ and $n_j=\ka+\ceil{\frac{\nu_c+j}2}$. 
The corresponding residue projects out the single integer coefficient $G_{\ka-k+j,\ka-k,\nu_c-1}$ only, 
corresponding to the first term of the sum \eq\nr{eq:kallen}, while the $\intB^{\ka,\ka,\nu_c}(d)$ result reduces to two terms. 
Inverting the first of \eq\nr{eq:G}, we can map $g_{a_1,a_2,a_3}=\frac12\,G_{a_2,a_3,a_1}$ onto this expression 
(which works for the case $a_2\le a_3$ for $j=a_2-a_3$, $\ka=a_3+1$ and $2\ka+\nu_c=3+a_1+2a_3$, from which follow $k=1$ 
and $n_j=2+\frac{a_1+a_2+a_3}2$; all the time remembering that $a_1+a_2+a_3$ is even), whence only the first of the two terms 
of the $\intB^{\ka,\ka,\nu_c}(d)$ result contributes since we probe it with $\nu_c+j$=odd here. 
The resulting formula can finally be symmetrized in its indices, leading to 
\ba\la{eq:gResult}
g_{a_1,a_2,a_3} &=& \frac{4^A\,A!}{a_1!\,a_2!\,a_3!\,\po{\frac12}{a_1-A}\,\po{\frac12}{a_2-A}\,\po{\frac12}{a_3-A}} 
\;=\;
 \frac{(-4)^A\,A!}{a_1!\,a_2!\,a_3!}\,\po{\tfrac12}{A-a_1}\,\po{\tfrac12}{A-a_2}\,\po{\tfrac12}{A-a_3}
\;,\quad
\ea
with integer $A=\frac{a_1+a_2+a_3}2$ as above, and where both versions can be seen to be 
equivalent by using the definition of the Pochhammer symbols in terms of Gamma functions (cf.\ \eq\nr{eq:defPoch}) and using elementary relations between those.
Additionally, Legendre's duplication formula for the Gamma function could be employed to rewrite Pochhammer symbols of positive half-integer values 
in terms of factorials as
$\po{k+\tfrac12}{n} = \frac{\Gamma(n+k+\frac12)}{\Gamma(k+\frac12)} \;=\; \frac{(2n+2k)!\,k!}{4^n\,(n+k)!\,(2k)!}$ (valid for positive $n$),
but we prefer to keep the forms given above to avoid distinguishing between positive and negative values of the integers $A-a_i$.

Most amazingly, even though we have just looked at a single coefficient in the $\Nu+j$=odd case, 
this fixes {\em all} coefficients $g$, hence also the $G_{a,b,c}$ which are needed for $\Nu+j$=even cases. 
Note also that \eq\nr{eq:gResult} actually works correctly for {\em all} integer values of the indices $a_i$: if any $a_i<0$, the corresponding inverse factorial $1/a_i!=1/\Gamma(a_i+1)=1/\Gamma(\setZ\le0)$ nullifies $g$; furthermore, if $a_1+a_2+a_3$ is odd, the inverse Pochhammer corresponding to the smallest $a_i$ does the same, $1/\po{\frac12}{{\min}\{a_i\}-A}=\Gamma(\frac12)/\Gamma(\frac{1+2{\min}\{a_i\}-a_1-a_2-a_3}2)=\Gamma(\frac12)/\Gamma(\setZ\le0)$.

As a check, the zero index cases \eq\nr{eq:gZero} follow immediately from \eq\nr{eq:gResult}. 
It also follows that for $(a+b+c)$ odd, the 3-$g$ combination $G_{a,b,c}$ can be written as a single $g$ (with integer $s\equiv\frac{1+a+b+c}2$)
\ba \la{eq:G0}
G_{a,b,c}\,\Big|_{(a+b+c)~{\rm odd}} &=& 
\frac{4^{s-\frac12}\,\Gamma(s)}{a!\,b!\,c!\,\po{\frac12}{a-s}\,\po{\frac12}{b-s}\,\po{\frac12}{1+c-s}}
\;=\; \frac{1+c}{1+a+b+c}\;g_{a,b,1+c} \;.
\ea

We close this section by further simplifying the previous section's final result. 
The coefficients $G$ that appear in \eq\nr{eq:kallen} and have been expressed in terms of single $g_{a,c,b}$ in \eqs\nr{eq:G0} and \nr{eq:G} 
for even and odd cases, respectively, can be represented in a unified way. 
Using that for $\Nu+j$=even (odd) we have $j=2n_j-\Nu$ ($j=2n_j-\Nu-1$) and slightly massaging the Pochhammer symbols by going back
to their definition in terms of Gamma functions (cf.\ \eq\nr{eq:defPoch}), we obtain
\ba
G_{\nu_a+j-k,\nu_b-k,\nu_c-1} &=& 
\frac{2\cdot 4^{n_j-k-1}\,(n_j-k-1)!}{(\nu_a+j-k)!\,(\nu_b-k)!\,(\nu_c-1)!\,\po{\frac12}{n_j-\nu_b-\nu_c}\po{\frac12}{n_j-j-\nu_a-\nu_c}\po{\frac12}{\nu_c-n_j+k}} 
\;.\qquad
\ea
This allows to rewrite \eq\nr{eq:kallen} as (all factorials good: $\nu_c>0 \Rightarrow n_j>k$; rest from summation limits)
\ba\la{eq:kallen3}
c^{(\Nu)}_{\nu_a,\nu_b;j}(d) &=& 
\frac{(-1)^{\Nu-n_j+1}\,\po{1-\frac{d}2}{n_j-j-1}}{2\,\po{\frac12}{n_j-\nu_b-\nu_c}\,\po{\frac12}{n_j-j-\nu_a-\nu_c}\,(\nu_c-1)!}
\times\nonumber\\&\times& 
\sum_{k={\rm max}(1+j,1)}^{{\rm min}(\nu_a+j,\nu_b)} 
\frac{\po{\frac{d}2-n_j+1}{k-1}\,(n_j-k-1)!}{\po{\frac{d+3}2\!-\!\Nu}{n_j-k} \po{\frac12}{\nu_c-n_j+k}\,(k\!-\!1)!\,(k\!-\!j\!-\!1)!\,(\nu_b\!-\!k)!\,(\nu_a\!-\!k\!+\!j)!}
\;,\qquad
\ea
where we recall that $n_j=\ceil{\frac{\Nu+j}2}$ and $\Nu=\nu_a+\nu_b+\nu_c$ here.

Using this explicit solution, one can test for symmetries among the rational functions $c^{(\Nu)}$, in addition to the ones given
already in \eq\nr{eq:csy}. Among all 58870 coefficients up to weight 30, we discovered an additional general symmetry 
\ba
c^{(\Nu)}_{\nu_a,\nu_b;j}(d) &=& c^{(\Nu)}_{\nu_b-j,\nu_a+j;j}(d)
\ea
that leaves $\nu_a+\nu_b$ unchanged, as well as two more particular ones 
\ba
c^{(\Nu)}_{1,\nu_b;j}(d) \;=\; (-1)^{\Nu-1+j}c^{(\Nu)}_{1,\Nu-1-\nu_b+j;j}(d) \;,\quad
c^{(\Nu)}_{\nu_a,1;j}(d) \;=\; (-1)^{\Nu-1-j}c^{(\Nu)}_{\Nu-\nu_a-1-j,1;j}(d) \quad
\ea
which, depending on the parity of $\Nu+j$, also relate coefficients with 
different overall signs. All symmetries leave the weight as well as $|j|$ unchanged. So only 14875 of those coefficients are independent 
functions. 

For example, the three relations between the coefficients that one can observe in \eq\nr{eq:exampleB123} are captured by 
these symmetry relations. In fact, of the 30 distinct rational coefficient functions $c^{(6)}_{a,b:j}(d)$ needed to parametrize all 
ten weight-6 integrals $\intB^{1,2,3}(d)$, $\intB^{1,1,4}(d)$, $\intB^{2,2,2}(d)$ (plus index permutations) according 
to \eq\nr{eq:newconj}, only 9 are independent (up to an overall sign; at larger weight, the fraction of such independent coefficients 
approaches 25\% from above), due to relations such as the six-fold symmetry 
$c^{(6)}_{2,3;2}=c^{(6)}_{3,2;-2}=c^{(6)}_{1,4;2}=c^{(6)}_{4,1;-2}=-c^{(6)}_{1,3;2}=-c^{(6)}_{3,1;-2}=-\frac{(d-2)(d-4)(d-6)}{32(d-9)}$. 
Note that some of these symmetries relate coefficients belonging to different integrals, such as to $\intB^{1,2,3}(d)$ and $\intB^{1,1,4}(d)$ 
in the example shown here. 

Note also that other simple relations, such as $c^{(6)}_{1,3;1}=2c^{(6)}_{1,2;1}$, cf.\ \eq\nr{eq:exampleB123}, 
are not yet captured, so it is not excluded that one can discover additional symmetry relations in \eq\nr{eq:kallen3}.
We do not explore this further here, since having the explicit solution at hand is fully sufficient.

%
\section{Application: three-point integrals for collinear momenta}
\la{se:applications}

Returning to the discussion of \se\ref{se:intro}, we note that for three-point functions (with external momenta $p_1$, $p_2$ and hence $p_3=-p_1-p_2$ for momentum conservation), the \kallen function $\lambda(p_1^2,p_2^2,p_3^2)$ vanishes in the collinear limit (i.e.\ for parallel external momenta: $p_1=x P$, $p_2=(1-x)P$ where $P=p_1+p_2=-p_3$ is the total momentum, and $x$ is the fraction thereof carried by the first external particle).

It is therefore natural to ask whether one can also expect a significant simplification of three-point functions in the collinear limit, analogous to the factorization property of ``mass-collinear'' vacuum integrals proven above.
In the one-loop case with massless internal lines the answer turns out to be affirmative, as we will demonstrate in this section. 

The key ingredient is an astonishing exact relation between 1-loop massless triangles 
\ba\la{eq:defJ}
\intJ^\nabc(d;p_1^2,p_2^2,p_3^2) &\equiv& 
\intd{q}
\frac{1}{[(q-p_2)^2]^{\nu_1}\,[(q+p_1)^2]^{\nu_2}\,[q^2]^{\nu_3}} \;,\quad
p_3^2=(p_1+p_2)^2
\ea
and the 2-loop massive tadpoles $\intB$ of \eq\nr{eq:defB1} (with three independent masses still), which has been discovered in~\cite{Davydychev:1995mq} 
and dubbed the ``magic'' connection therein.\footnote{In order to translate between \cite{Davydychev:1995mq} and our Euclidean notation, 
we note that $J(d;\nu_1,\nu_2,\nu_3|-p_1^2,-p_2^2,-p_3^2)=(-1)^\Nu\,i\,\pi^{d/2}\,L^{-d}\,T^\nabc(d;p_1^2,p_2^2,p_3^2)$. See also footnotes \ref{foot:1} and \ref{ft:L}.} 
The original derivation is based on the similarity of general results for these integrals, which (in both cases) can be expressed in terms of Appell's $F_4$ functions~\cite{Boos:1987bg,Boos:1990rg}, as well on the similar structure of the respective recurrence relations~\cite{Davydychev:1992xr}. 

In the following two subsections, we first recall the derivation of the ``magic'' relation via Feynman parameter representations, and then apply the factorization property \eq\nr{eq:newconj2} to obtain the abovementioned simplifications for the collinear-momenta three-point case.

%
\subsection{Re-deriving the ``magic'' connection}

The Feynman parameter representations for our two types of three-propagator integrals are both three-fold and read
\ba\la{eq:Bfy}
\frac{\intB^\nabc(d;m_1^2,m_2^2,m_3^2)}{L^{2d}\, \Gamma(\Nu-d)} &=& 
\int_{x_i\ge0}\Big(\prod_{j=1}^3\frac{{\rm d}x_j\,x_j^{\nu_j}}{x_j\,\Gamma(\nu_j)}\Big)\,
\frac{\delta(x_1+x_2+x_3-1)\;(x_1x_2+x_1x_3+x_2x_3)^{\frac{d}2}}{(x_1m_1^2+x_2m_2^2+x_3m_3^2)^{\Nu-d}} \;,\qquad\\
\la{eq:Jfy}
\frac{\intJ^\nabc(d;p_1^2,p_2^2,p_3^2)}{L^{d}\, \Gamma(\Nu-\frac{d}2)} &=& 
\int_{x_i\ge0}\Big(\prod_{j=1}^3\frac{{\rm d}x_j\,x_j^{\nu_j}}{x_j\,\Gamma(\nu_j)}\Big)\,
\frac{\delta(x_1+x_2+x_3-1)}{(x_1x_2p_3^2+x_1x_3p_2^2+x_2x_3p_1^2)^{\Nu-\frac{d}2}} \;,
\ea
where we recall that $L=(4\pi)^{-\frac12}$ is due to our integral normalization, and $\Nu=\nu_1+\nu_2+\nu_3$.

To fully expose the similarity of the two expressions, one can perform a conformal-type integration variable transformation $x_i=\frac{z_1z_2z_3(z_1+z_2+z_3)}{z_i(z_1z_2+z_2z_3+z_3z_1)}$ that is chosen such that the argument of the delta function remains unchanged \cite{Scharf:1993ds,Davydychev:1995mq}. After considering the homogeneity of the integrand's denominator, many factors cancel against the Jacobian of the transformation. The Feynman parameter representation \eq\nr{eq:Bfy} of the massive two-loop vacuum diagram then takes the equivalent form
\ba
\frac{\intB^\nabc(d;m_1^2,m_2^2,m_3^2)}{L^{2d}\, \Gamma(\Nu-d)} &=& 
\int_{z_i\ge0}\Big(\prod_{j=1}^3\frac{{\rm d}z_j\,z_j^{\Nu-\frac{d}2-\nu_j}}{z_j\,\Gamma(\nu_j)}\Big)\,
\frac{\delta(z_1+z_2+z_3-1)}{(z_1z_2m_3^2+z_1z_3m_2^2+z_2z_3m_1^2)^{\Nu-d}} \;.
\ea
Comparing this with \eq\nr{eq:Jfy} allows to read off the exact relation
\ba\la{eq:BJrel}
\intB^\nabc(d;m_1^2,m_2^2,m_3^2) &=& L^{3d-2\Nu}\,
\Big(\prod_{j=1}^3\frac{\Gamma(\tilde\nu_j)}{\Gamma(\nu_j)}\Big)\,
\intJ^{\tilde\nu_1,\tilde\nu_2,\tilde\nu_3}(2\Nu-d;m_1^2,m_2^2,m_3^2) \;,
\ea
where the right-hand side contains the massless triangle integral, with $d$\/-dependent indices that we abbreviate 
as $\tilde\nu_j\equiv\Nu-\frac{d}2-\nu_j$, and evaluated at a shifted dimension. This latter integral can still be simplified, as explained in the following.

Massless loop integrals sometimes allow for simple evaluations. Two such examples are
\ba\la{eq:Gint}
&&\intG^{\nu_1,\nu_2}(d;p^2) \equiv  
\intd{k}
\frac{1}{[k^2]^{\nu_1}[(k+p)^2]^{\nu_2}} \;=\;
L^d\,\frac{\Gamma(\nu_1+\nu_2-\frac{d}2)\,\Gamma(\frac{d}2-\nu_1)\,\Gamma(\frac{d}2-\nu_2)}{\Gamma(d-\nu_1-\nu_2)\,\Gamma(\nu_1)\,\Gamma(\nu_2)\,[p^2]^{\nu_1+\nu_2-\frac{d}2}} \;,\\
\la{eq:uniq}
&&\intJ^\nabc(d;p_1^2,p_2^2,p_3^2) \stackrel{d=\Nu}=
\prod_{j=1}^3\frac{\Gamma(\frac{\Nu}2-\nu_j)}{L^{-\nu_j}\,\Gamma(\nu_j)\,[p_j^2]^{\frac{d}2-\nu_j}}
\;.
\ea
The first of these is the trivial massless one-loop two-point (or $p$\/-) integral that has been known for a long time (see e.g.\ \cite{Chetyrkin:1980pr}). 
The formula is generic in the sense that the propagator powers $\nu_i$ do not need to be integers, but can for example also depend on the dimension $d$. 
It allows to recursively integrate large classes of multi-loop $p$\/-integrals (see e.g. \cite{Chetyrkin:1980pr,Baikov:2010hf,Lee:2011jt,Georgoudis:2018olj,Georgoudis:2021onj} 
for 3- to 5-loop examples). 
Equation \nr{eq:uniq} has been known at least since the 1980's as well \cite{Vasiliev:1981dg} in connection with so-called ``uniqueness'' 
methods \cite{Usyukina:1983gj,Kazakov:1983ns}. 
It has many derivations (see also \cite{Gracey:1992ew,Gracey:2013sz,Kotikov:2018wxe}), and can also be obtained by integrating \eq\nr{eq:Jfy} above.

One of the nice tricks of the uniqueness methods is then to use the two relations in combination, in order to systematically shift indices of 
massless triangle integrals: starting from a triangle with index sum $\Nu\neq d$, in order to enforce $\Nu= d$ one splits one of the lines 
into two using \eq\nr{eq:Gint} in reverse, at the cost of an additional loop integral; then \eq\nr{eq:uniq} is used on the enforced unique triangle, 
reducing the loop number again and producing an external leg at the vertex that is non-adjacent to the line that had been split. 
For our massless triangles $\intJ^\nabc$ of \eq\nr{eq:defJ}, one move of the uniqueness game is played as follows: pick one of the three 
propagators (say, the third), identically rewrite its corresponding index ($\nu_3=[d-\nu_1-\nu_2]+[\Nu-\frac{d}2]-\frac{d}2$) 
and use \eq\nr{eq:Gint} in reverse ($[q^2]^{-\nu_3}\sim \intG^{d-\nu_1-\nu_2,\Nu-\frac{d}2}(d;q^2)$ up to Gamma factors), 
whence the first of the two propagators in the one-loop bubble $\intG$ forms a unique triangle with the remaining two propagators 
of $\intJ$ allowing to use \eq\nr{eq:uniq}, which results in 
\ba
\intJ^\nabc(d;p_1^2,p_2^2,p_3^2) &=& 
\frac{\Gamma(\Nu-\frac{d}2)}{\Gamma(d-\Nu)}\,
\Big(\prod_{j=1}^3\frac{\Gamma(\frac{d}2-\nu_j)}{\Gamma(\nu_j)}\Big)\,
\frac{\intJ^{\frac{d}2-\nu_2,\frac{d}2-\nu_1,\Nu-\frac{d}2}(d;p_1^2,p_2^2,p_3^2)}{[p_3^2]^{\Nu-\nu_3-\frac{d}2}} \;.
\ea
Repeating this procedure for each of the three lines, one obtains
\ba\la{eq:Junique}
\intJ^\nabc(d;p_1^2,p_2^2,p_3^2) &=&
\frac{\Gamma(\Nu-\frac{d}2)}{\Gamma(d-\Nu)}\,\Big(\prod_{j=1}^3\frac{\Gamma(\frac{d}2-\nu_j)}{[p_j^2]^{\Nu-\nu_j-\frac{d}2}\,\Gamma(\nu_j)}\Big) 
\intJ^{\frac{d}2-\nu_1,\frac{d}2-\nu_2,\frac{d}2-\nu_3}(d;p_1^2,p_2^2,p_3^2) \:.\quad
\ea 
The magic connection of \cite{Davydychev:1995mq} then follows from combining \eq\nr{eq:Junique} with \eq\nr{eq:BJrel}, resulting in
\ba\la{eq:magicRel}
\intJ^\nabc(d;p_1^2,p_2^2,p_3^2) &=& 
L^{3d-4\Nu}\,
\frac{\Gamma(\Nu-\frac{d}2)}{\Gamma(d-\Nu)}\,
\frac{\intB^\nabc(2\Nu-d;p_1^2,p_2^2,p_3^2)}{\prod_{k=1}^3[p_k^2]^{\Nu-\nu_k-\frac{d}2}} \;,
\ea
where we recall that $\Nu=\nu_1+\nu_2+\nu_3$ is the index sum, and $L=(4\pi)^{-\frac12}$ is a normalization factor.

%
\subsection{Mapping collinear three-point integrals onto two-point functions}
\la{se:mapping}

Having established the connection \eq\nr{eq:magicRel} in our notation, we can now restrict both sides of the equation to the special case of vanishing \kallen function. 
As already discussed above, $\lambda(p_1^2,p_2^2,p_3^2)=0$ implies collinear momenta $p_1\parallel p_2$.
Imposing these kinematic restrictions, all we have learned about the massive vacuum integrals $\intB$ can be directly applied to the massless triangles $\intJ$.
In particular, applying \eq\nr{eq:newconj2} (with $d\rightarrow2\Nu-d$ and $m_i\rightarrow(p_i^2)^{\frac12}$; and inserting into each sum 
one factor like $1=\frac{\intG^{\nu_1,\nu_2}(d;p_3^2)}{\intG^{\nu_1,\nu_2}(d;1)}\,[p_3^2]^{\nu_1+\nu_2-\frac{d}2}$ etc., 
in order to eliminate $d$ from explicit powers of the invariants $p_i^2$), we can immediately reduce one of the positive propagator 
powers $\nu_i$ of three-point integrals $\intJ^\nabc$ to zero, thus obtaining a linear combination of the trivial massless two-point 
functions $\intG$ of \eq\nr{eq:Gint}, since $\intJ^{\nu_1,\nu_2,0}(d;p_1^2,p_2^2,p_3^2) = \intG^{\nu_1,\nu_2}(d;p_3^2)$, with similar relations for the other two cases. In full structural analogy to \eq\nr{eq:newconj}, the reduction reads
\ba
\intJ^\nabc(d;p_1^2,p_2^2,p_3^2) &\stackrel{p_1\parallel p_2}=& 
\frac{\intG^{\nu_1,\nu_2}(d;p_3^2)}{(p_1^2)^{\frac{\Nu}2-\nu_1}\,(p_2^2)^{\frac{\Nu}2-\nu_2}}\,
\sum_{j=1-\nu_1}^{\nu_2-1}(-1)^\Nu\, \barc^{(\Nu)}_{\nu_1,\nu_2;j}(d)\, \Big(\frac{p_1^2}{p_2^2}\Big)^{\frac{j}2}
\nonumber\\&+&
\frac{\intG^{\nu_1,\nu_3}(d;p_2^2)}{(p_1^2)^{\frac{\Nu}2-\nu_1}\,(p_3^2)^{\frac{\Nu}2-\nu_3}}\,
\sum_{j=1-\nu_1}^{\nu_3-1}(-1)^j\, \barc^{(\Nu)}_{\nu_1,\nu_3;j}(d)\, \Big(\frac{p_1^2}{p_3^2}\Big)^{\frac{j}2}
\nonumber\\&+&
\frac{\intG^{\nu_2,\nu_3}(d;p_1^2)}{(p_2^2)^{\frac{\Nu}2-\nu_2}\,(p_3^2)^{\frac{\Nu}2-\nu_3}}\,
\sum_{j=1-\nu_2}^{\nu_3-1}(-1)^j\, \barc^{(\Nu)}_{\nu_2,\nu_3;j}(d)\, \Big(\frac{p_2^2}{p_3^2}\Big)^{\frac{j}2}
\;,\\ \mbox{with}\qquad
\barc^{(\Nu)}_{\nu_1,\nu_2;j}(d) &=& \frac{\Gamma(\Nu-\frac{d}2)\,\Gamma^2(1-\Nu+\frac{d}2)}{\Gamma(d-\Nu)\,L^{-d}\,\intG^{\nu_1,\nu_2}(d;1)}\,
c^{(\Nu)}_{\nu_1,\nu_2;j}(2\Nu-d) \;,
\ea
and coefficient $c^{(\Nu)}_{\nu_1,\nu_2;j}$ from \eq\nr{eq:kallen3}.

\begin{figure}
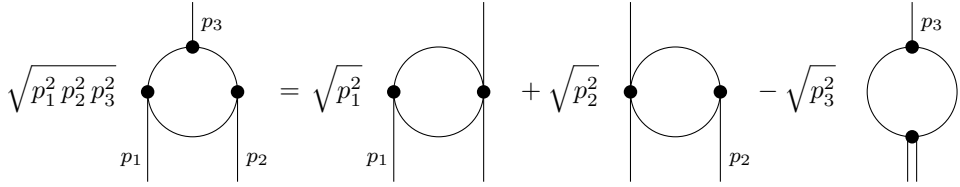

\setlength{\unitlength}{1.7pt}
\ba \nonumber
\sqrt{p_1^2\,p_2^2\,p_3^2}\Ja &=& \sqrt{p_1^2}\Jb + \sqrt{p_2^2}\Jc - \sqrt{p_3^2}\Jd
\ea
\caption{\la{fig:magic}Structure of the reduction of massless one-loop three-point functions with collinear momenta to one-loop two-point functions, 
such as for the specific example given in \eq\nr{eq:J111}. Here, all internal lines correspond to single-power massless progagators $1/[k^2]$.
We label the external momenta by $p_i$, implying collinearity $p_1\parallel p_2$, which we indicate by drawing the external lines parallel.}
\end{figure}

For example (see also \fig\ref{fig:magic}), the massless one-loop three-point function with collinear momenta $p_1\parallel p_2$ and unit propagator powers reduces to one-loop two-point functions as ($\barc^{(3)}_{1,1;0}(d)=1$)
\ba\la{eq:J111}
\intJ^{1,1,1}(d;p_1^2,p_2^2,p_3^2) &\stackrel{p_1\parallel p_2}=& 
\frac{\intG^{1,1}(d;p_1^2)}{(p_2^2\,p_3^2)^{\frac12}} 
+\frac{\intG^{1,1}(d;p_2^2)}{(p_3^2\,p_1^2)^{\frac12}} 
-\frac{\intG^{1,1}(d;p_3^2)}{(p_1^2\,p_2^2)^{\frac12}} \;.
\ea

Given that $p$\/-integrals constitute a well-studied class of multi-loop Feynman integrals 
(see e.g.\ \cite{Chetyrkin:1980pr}, \cite{Baikov:2010hf,Lee:2011jt} and \cite{Georgoudis:2018olj} for three-, four- and five-loop work; 
see also the review \cite{Kotikov:2018wxe}), a natural question would be to ask whether the mapping from $\intJ$ to $\intG$ can be 
generalized to higher loops. We leave this as an interesting open question for the future.

%
\section{Summary and Outlook}
\la{se:summary}

Considering the effort that went into establishing the partial results 
\eq\nr{eq:BaacM} (closed-form solution of the particular-index case $\intB^{\ka,\ka,\nu}(d)$), 
\eq\nr{eq:newconj2} (mass dependence of the general-index case $\intB^{\nabc}(d)$) and 
\eq\nr{eq:kallen} (functional $d$\/-dependence of the general-index case) 
by largely independent methods, 
it has been extremely gratifying to see all three ends meet in \se\ref{se:4.2}, and be able to extract an explicit result for the integral $\intB^\nabc_\mabc(d)$ of \eq\nr{eq:defB2}.
From \eq\nr{eq:newconj2}, \eq\nr{eq:masters} and \eq\nr{eq:kallen3}, our general factorization formula takes the form of a finite double-sum
\ba \la{eq:6.1}
\frac{\intB^\nabc_\mabc(d)}{L^{2d}\,\Gamma^2(1-\tfrac{d}2)} &=& 
\sum_{j=1-\nu_1}^{\nu_3-1} \frac{c^{(\Nu)}_{\nu_1,\nu_3;j}(d)}{(m_1m_3)^{\Nu-d}} \, \Big(-\frac{m_1}{m_3}\Big)^j 
+\sum_{j=1-\nu_2}^{\nu_3-1} \frac{c^{(\Nu)}_{\nu_2,\nu_3;j}(d)}{(m_2m_3)^{\Nu-d}} \, \Big(-\frac{m_2}{m_3}\Big)^j 
\nonumber\\&+&
\sum_{j=1-\nu_1}^{\nu_2-1} (-1)^\nu \, \frac{c^{(\Nu)}_{\nu_1,\nu_2;j}(d)}{(m_1m_2)^{\Nu-d}} \, \Big(\frac{m_1}{m_2}\Big)^j 
\,\Bigg|_{\scriptsize\begin{array}{l} m_3=m_1+m_2 \\ \Nu=\nu_1+\nu_2+\nu_3 \end{array}} \;,\\
\la{eq:6.2}
c^{(\Nu)}_{\nu_a,\nu_b;j}(d) &=& 
\frac{(-1)^{\Nu-n_j+1}\,\po{1-\frac{d}2}{n_j-j-1}}{2\,\po{\frac12}{n_j-\nu_b-\nu_c}\,\po{\frac12}{n_j-j-\nu_a-\nu_c}\,(\nu_c-1)!}
\times\nonumber\\&\times& 
\sum_{k={\rm max}(1+j,1)}^{{\rm min}(\nu_a+j,\nu_b)} 
\frac{\po{\frac{d}2-n_j+1}{k-1}\,(n_j-k-1)!}{\po{\frac{d+3}2\!-\!\Nu}{n_j-k} \po{\frac12}{\nu_c-n_j+k}\,(k\!-\!1)!\,(k\!-\!j\!-\!1)!\,(\nu_b\!-\!k)!\,(\nu_a\!-\!k\!+\!j)!}
\qquad
\ea
with Pochhammer symbols $(a)_\nu\equiv\frac{\Gamma(a+\nu)}{\Gamma(a)}$ and integers $n_j=\ceil{\frac{\Nu+j}2}$. 
The normalization factor on the left-hand side of \eq\nr{eq:6.1} is the square of a massive one-loop tadpole, cf.\ eq\nr{eq:masters}. 

We have gone full circle, guided by the \kallen function of \eq\nr{eq:defKallen}. 
First, its zeros provided us with a linear propagator mass relation. 
Then, at these \kallen zeros, IBP reduction relations for our massive two-loop vacuum integral simplified to a degree that allowed to 
extract closed-form analytic results from some special-mass and special-index cases
such as $\intB^\nabc_{0,m,m}(d)$ of \se\ref{se:B0mm} and $\intB^{\ka,\ka,\nu}(d)$ of \se\ref{se:Baac}, as well as to prove the general mass dependence shown in \eq\nr{eq:6.1}. 
Next, and perhaps surprisingly, the \kallen function resurfaced in operator form at the core of the powerful two-step IBP reduction relation \eq\nr{eq:AD82}.
Finally, it not only governed the generating function for the combinatorics at the core of the coefficients $c^{(\Nu)}(d)$ as described in \se\ref{se:4.1},
but also allowed for a complete solution of that combinatorial problem, leading to the explicit form \eq\nr{eq:6.2} of those coefficient functions.

Looking back, we note that in the case of $\intB^{\ka,\ka,\nu}(d)$ we have solved a 
two-dimensional linear homogeneous recurrence with variable coefficients that underlies \eq\nr{eq:T94b} (see \app\ref{app:guts});
while for deriving the coefficients in the full solution for $\intB^\nabc(d)$, after having proven its generic mass-structure, 
we identified and solved a three-dimensional linear homogeneous recurrence with constant coefficients that underlies
\eq\nr{eq:AD82}, as explained in \se\ref{se:4.1}.
Both of these solutions of specific multidimensional recurrences seem currently out of reach for 
general summation techniques, which mostly rely on powerful symbolic summation tools that have been developed for the 
univariate case (see, e.g.\  \cite{Schneider:2013zna,Bluemlein:2022eym} and references therein).

A first application of our factorization formula has been to massless collinear three-point functions in \se\ref{se:applications},
where it had allowed to give a general decomposition of the triangle function into massless propagator integrals.
This is in close analogy to the well-known triangle rule \cite{Chetyrkin:1981qh,Tkachov:1981wb} 
(see also the related diamond rule \cite{Ruijl:2015aca}) that comes up when reducing
one-loop (sub-)diagrams via IBP relations, which had been solved in \cite{Tkachov:1984xk}. 
These closed forms for general-index cases are solutions in the sense that they express the result as a non-recursive linear 
combinations over simpler (lower-loop or lower-point) structures. 
The advantage of such explicit solutions is that fewer terms are generated (than in IBP recursions), 
and that spurious poles at intermediate steps of a recursion can be avoided. 

Another straightforward, albeit rather mathematical, application would be to extract explicit results for 
certain types of Bessel moments. In $d$\/-dimensional coordinate space, massive propagators correspond 
to Bessel functions, such that after the trivial angular integration corresponding to a two-point function 
one is left with a one-dimensional radial integral representation for
the vacuum integral of \eq\nr{eq:defB1}, 
$\intB^\nabc\propto \int_0^\infty {\rm d}x\,x^{\Nu-d/2-1}\,K_{d/2-\nu_1}(m_1x)\,K_{d/2-\nu_2}(m_2x)\,K_{d/2-\nu_3}(m_3x)$ 
(see \cite{Mendels:1978wc,Berends:1993ee,Groote:1998wy} for 
related treatments of sunset-type integrals).
Setting $m_3=m_1+m_2$ and comparing with \eq\nr{eq:6.1} then allows to find explicit results for such
triple-Bessel moments with general indices.

One might be tempted to further simplify the above result and solve one of the two remaining finite sums, perhaps after commuting them. 
We have not been able to do this ourselves, but regard our final result as highly useful in the present form, as for most practical 
cases the indices $\nu_i$ have rather small integer values, such that only very few terms contribute to the double sum.
Other interesting generalizations would be to derive compact results for other special-index cases other than $\intB^{\ka,\ka,\nu}(d)$,
such as e.g.\ for $\intB^{1,\nu_b,\nu_c}(d)$, or to look at analogous simplifications (for phenomenologically
relevant kinematic constraints such as the linear mass-relations exploited here) of higher-loop IBP relations and attempt progress
on their generic solutions. On the more formal side, it might be possible to establish a connection between the specific two-loop factorization 
observed here and Baikov's (ir)reducibility criterion \cite{Baikov:2005nv}.

We close by noting that the mass structure of our integral is similar to what appears within finite temperature field theory, and this is no coincidence; 
as already mentioned in \se\ref{se:intro}, it had in fact been one of the main motivations that led us into the present work. 
Indeed, in the so-called imaginary time formalism, 
the temporal components of the momentum four-vectors obey a linear relation at each vertex, akin to our \eq\nr{eq:massRel}. 
The detailed investigation of such finite-temperature sum-integrals in the light of our new results is well beyond the scope of the present work, 
however, and we leave it for the future \cite{sumint2}.

%
\acknowledgments

We wish to thank Igor Kondrashuk for discussions at early stages of this work,
and Pablo Navarrete for a critical reading of parts of the manuscript.
A.D.\ would like to thank the Universidad del Bío-Bío for hospitality during stages of this work, 
and acknowledges partial support by CONICYT PCI/MEC 80180071 and FONDECYT project 1191073.
Y.S.\ acknowledges support from FONDECYT projects 1151281 and 1191073.

%
\appendix

%
\section{Derivation of \eq\nr{eq:418}}
\la{se:418proof}

\begin{figure}
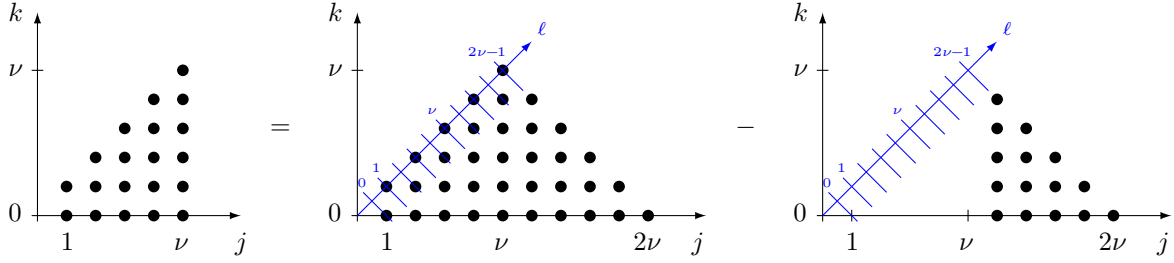

\begin{center}
\setlength{\unitlength}{1.1pt}
\doubleSumSplit
\caption{\label{figC} Rewriting the double sum of \eq\nr{eq:sumC} by adding and subtracting a suitable $(j,k)$\/-region. On the right-hand side, we have also indicated the new summation variable $\ell$ that is used instead of $j$ in \eq\nr{eq:sumCC}.}
\end{center}
\end{figure}

To show how \eq\nr{eq:418} follows from \eq\nr{eq:417}, we denote the mass ratio as $x\equiv\frac{m_1}{m_3}$ 
such that \eq\nr{eq:417} reads (we suppress the overall $j$\/-independent factor $\frac{\intB^{1,0,1}(d)}{(m_1\,m_3)^\nu}$)
\ba
S&\equiv&\sum_{j=1}^\nu p_{\nu,j}\,\frac{[(d-j-1)\,x^{j-1}+(2j-d)\,x^j]}{(1-x)^{\nu+1-j}} \;,\\
\mbox{with}&& 
p_{\nu,j} \;\equiv\; \frac{\po{2-d+j}{\nu-j}}{\po{\frac{3-d}2+j}{\nu-j}}\,\frac{\po{1-\frac{d}2}{j-1}\,4^{-\nu+j-1}}{\Gamma(j)\,(j+\frac{1-d}2)} \;.
\ea
Replace $x\rightarrow[1-(1-x)]$ in the numerator and use the binomial formula to expand as
\ba
x^{j-1} = \sum_{k=0}^j \bin{j}{k}\frac{j-k}{j}\,(-1)^k(1-x)^k \;,\qquad
x^j = \sum_{k=0}^j \bin{j}{k}(-1)^k(1-x)^k \;,
\ea
such that 
\ba\la{eq:sumC}
S &=& \sum_{j=1}^\nu \sum_{k=0}^j p_{\nu,j}\,q_{k,j}\,(1-x)^{k+j-\nu-1} \;,\\
\mbox{with}&&
q_{k,j} \;\equiv\; \bin{j}{k}(-1)^k[(d-j-1)\tfrac{j-k}{j}+(2j-d)] \;.
\ea
Rewrite the double sum by adding and subtracting a number of $(j,k)$\/-points (see \fig\ref{figC}) and change the summation variable $j$ for $\ell\equiv j+k-1$, to obtain
\ba\la{eq:sumCC}
S &=& \Bigg\{ \sum_{\ell=0}^{2\nu-1}\sum_{k=0}^{\ceil{\ell/2}} - \sum_{\ell=\nu}^{2\nu-1}\sum_{k=0}^{\ell-\nu}\Bigg\}
p_{\nu,1+\ell-k}\,q_{k,1+\ell-k}\,(1-x)^{\ell-\nu} \;.
\ea
In the second double sum, we have $\ell\ge \nu$, such that the binomial can be expanded as 
$(1-x)^{\ell-\nu}=\sum_{j=0}^{\ell-\nu}\bin{\ell-\nu}{j}(-x)^j$, and after trading $\ell$ for $L\equiv\ell-\nu$ and exchanging $L$\/- and $j$\/-sums, we get
\ba
S &=& \sum_{\ell=0}^{2\nu-1} (1-x)^{\ell-\nu}\,r_{\nu,\ell} - \sum_{j=0}^{\nu-1}x^j\,s_{\nu,j} \;,\\
\mbox{with}&&
r_{\nu,\ell} \equiv \sum_{k=0}^{\ceil{\ell/2}} p_{\nu,1+\ell-k}\,q_{k,1+\ell-k} \;,\quad
s_{\nu,j} \equiv  (-1)^j \sum_{L=j}^{\nu-1}\sum_{k=0}^{L} \bin{L}{k}\,p_{\nu,1+\nu+L-k}\,q_{k,1+\nu+L-k} \;.\quad
\ea
Both sums can be evaluated immediately, to give
\ba
r_{\nu,\ell} &=& 0 \;,\\
s_{\nu,j} &=& \frac{\po{1-\frac{d}2}{\nu}}{j!}\,\sum_{\ell=j}^{\nu-1} \frac{(-1)^{\ell-j}/(2-d+\nu+\ell)}{(\ell-j)!\,\Gamma(\nu-\ell)}
\;=\; \frac{\po{1-\frac{d}2}{\nu}}{j!\,\po{2-d+\nu+j}{\nu-j}} \;.
\ea
Reinstalling the suppressed prefactor and setting $x=\frac{m_1}{m_3}$, we finally arrive at
\ba
\mbox{\eq\nr{eq:417}} &=& 
\frac{\intB^{1,0,1}(d)}{(m_1\,m_3)^\nu}\,S
\;=\; - \frac{\intB^{1,0,1}(d)}{(m_1\,m_3)^\nu} \sum_{j=0}^{\nu-1} \frac{\po{1-\frac{d}2}{\nu}}{j!\,\po{2-d+\nu+j}{\nu-j}}  \Big(\frac{m_1}{m_3}\Big)^j \;,
\ea
which is equivalent to (and slightly simpler than) \eq\nr{eq:418}.

%
\section{Exposing and solving the recurrence at the core of \eq\nr{eq:T94b}}
\la{app:guts}

Recurrence relations (or difference equations) are discrete equations that determine the elements of a sequence over the integers, 
given some initial conditions. Solving a recurrence relation amounts to deriving an explicit solution for this sequence in {\em non-recursive} form. 

In the case of one-dimensional linear recurrences for example, a sequence $f_n$ is determined for all $n\in\setZ$ from an 
$r$\/-th order recurrence relation $\sum_{k=0}^r a^{(k)} f_{n+k}=b$, given $r$ initial conditions $f_{n_i}$. 
In analogy to differential equations, the relation is called homogeneous when $b=0$, and the coefficients can be constant or variable 
(i.e.\ depend on the index $n$). There are solution strategies for various such types of recurrences, typically involving generating functions
and the roots of the characteristic polynomial.
A prominent example is the second-order homogeneous recurrence with constant coefficients $f_n=a\,f_{n-1}+b\,f_{n-2}$, 
with boundary conditions $f_0$ and $f_1$, whose special case $\{f_0,f_1,a,b\}=\{0,1,1,1\}$ generates the Fibonacci sequence.
The general linear first-order non-homogeneous recurrence relation with variable coefficients 
\ba\la{eq:C35}
f_{n+1} \;=\; a _n\,f_n+b_n 
\ea
can be solved in terms of an initial condition $f_{n_0}$ as
\ba\la{eq:C36}
f_n 
\;=\; \Big(\prod_{k=n_0}^{n-1} a_k\Big)\, f_{n_0} + \sum_{j=n_0}^{n-1} b_j\, \Big(\prod_{k=j+1}^{n-1} a_k\Big) \;.
\ea
If the coefficient $a_n$ vanishes for some integers $n=n_i$, the above form of the solution is useful for determining $f_n$ for all $n\le{\rm min}\{n_i\}$, while the term proportional to the initial condition $f_{n_0}$ is annihilated by its prefactor when determining $f_n$ for $n>{\rm min}\{n_i\}$. In the latter case, the relation itself enforces initial values as $f_{n_i+1}=b_{n_i}$, such that it is sufficient to set $f_n=\sum_{j=n_i}^{n-1} b_j\, \Big(\prod_{k=j+1}^{n-1} a_k\Big)$ for $n>n_i$.

For multi-dimensional linear recurrences, which define multi-dimensional arrays of values, much less is known. 
A well-known example are the binomial coefficients, defined by the recurrence (two-dimensional, linear, homogeneous, constant coefficients)
\ba\la{eq:Binomial}
b_{n,k}=b_{n-1,k}+b_{n-1,k-1} \;,\quad b_{0,k}=\delta_k \;,
\ea
with solution $b_{n,k}=\bin{n}{k}$ and bivariate generating function $g(x,y)\equiv\sum_{i,j}b_{i,j}x^i y^j=\frac1{1-x-xy}$. 
Another simple example are the Stirling numbers of second kind, defined by the recurrence 
(two-dimensional, linear, homogeneous, variable coefficients)
\ba\la{eq:Stirling}
S_{n,k}=k\,S_{n-1,k}+S_{n-1,k-1} \;,\quad S_{0,k}=\delta_k \;,
\ea
that is solved by the one-dimensional sum $S_{n,k}=\sum_{j=0}^k\frac{(-1)^{k-j}\,j^{n-1}}{(j-1)!\,(k-j)!}$, while no closed form of the generating function is known.

Looking now at the structure of the recurrence \eq\nr{eq:T94b}, it is clear that integrals $\intB^{\ka,\ka,\nu}$ 
can be reduced to a linear combination of similar integrals at $a=1$, but different values of $\nu$, as
\ba\la{eq:ansatz}
\intB^{1+N,1+N,\nu}(d) &=& \sum_{k=0}^N \Xc_{N,k}(d,\nu)\, (m_3^2\,\upC)^k (\subDD)^N \intB^{1,1,\nu}(d) \;,
\ea
where the $\Xc_{N,k}(d,\nu)$ are coefficient functions that need to be determined.

At $N=0$, this Ansatz reduces to $\intB^{1,1,\nu}(d)=\Xc_{0,0}(d,\nu)\,\intB^{1,1,\nu}(d)$, fixing $\Xc_{0,0}(d,\nu)\stackrel!=1$.

Applying the recurrence \eq\nr{eq:T94b} at $\nu_1=\nu_2=N$, using the Ansatz \eq\nr{eq:ansatz} on both sides, 
and comparing coefficients of $(m_3^2\,\upC)^j (\subDD)^n \intB^{1,1,\nu}(d)$ leads to a recurrence for $\Xc$
\ba
(d-2)n^2\,\Xc_{n,j}(d,\nu) &=& (1-\delta_{j,n})(d-2-2\nu)\,\Xc_{n-1,j}(d-2,\nu) +(1-\delta_{j,0})\,2\nu \,\Xc_{n-1,j-1}(d-2,\nu+1) \;.\nonumber
\ea

This looks impenetrable, as it involves values of the unknown coefficient functions $\Xc$ with all four parameters changing. 
Note however that pairs of values are related: the recurrence leaves $d-2n\equiv2c_1$ and $\nu+j\equiv c_2$ invariant. 
We hence set $d=2c_1+2n$ and $\nu=c_2-j$, which exposes the much simpler, but completely equivalent, two-parameter recurrence
\ba
(c_1-1+n)n^2 \,\Xc_{n,j} &=& (1-\delta_{j,n})(c_1-1-c_2+j+n)\,\Xc_{n-1,j} + (1-\delta_{j,0})(c_2-j)\,\Xc_{n-1,j-1} 
\ea
where we have dropped the common function arguments $(c_1,c_2)$.
The values of $c_1$ and $c_2$ are then substituted back only {\em after} solving the equation.
To clean up the equation further, we can remove the factor $a_n\equiv(c_1-1+n)n^2$
from the left-hand side 
by defining
\ba
\bar{\Xc}_{n,j} &\equiv& (-1)^{n+j} \Big(\prod_{j=1}^n a_j\Big)\,\Xc_{n,j} \;=\; (-1)^{n+j} \po{c_1}{n}\,[n!]^2\,\Xc_{n,j} \;,
\ea
where the sign has been introduced for convenience. The recurrence is now (with $c_3=1+c_2-c_1$)
\ba
\bar{\Xc}_{n,j} &=& (1-\delta_{j,n})(c_3-j-n)\bar{\Xc}_{n-1,j} + (1-\delta_{j,0})(c_2-j)\bar{\Xc}_{n-1,j-1} \;,\quad \bar{\Xc}_{0,0}=1 \;.
\ea
The function $\bar{\Xc}_{n,j}$ takes non-zero values only in the wedge $0\le j \le n$.
One can drop the $(1-\delta)$ factors by using a generalized boundary condition at $n=0$ for all $j\in\setZ$
\ba\la{eq:guts}
\bar{\Xc}_{n,j} = (c_3-j-n)\,\bar{\Xc}_{n-1,j} + (c_2-j)\,\bar{\Xc}_{n-1,j-1} \;,\quad \bar{\Xc}_{0,j}=\delta_j \;.
\ea

Undeterred by the fact that \eq\nr{eq:guts} looks very similar to -- but is slightly more complicated than -- the Stirling 
recurrence \eq\nr{eq:Stirling} (for which no closed-form solution exists), we also note its structural similarity to the 
binomial recurrence \eq\nr{eq:Binomial} which motivates us to try the Ansatz $\bar{\Xc}_{n,j}=\bin{n}{j}\,\tilde{\Xc}_{n,j}$. 
This takes care of the generalized boundary conditions, since $\bin{0}{j}=\delta_j$, resulting in the new recurrence 
(after dividing by $\frac1n\bin{n}{j}=\frac1{n-j}\bin{n-1}{j}=\frac1j\bin{n-1}{j-1}$)
\ba
n\,\tilde{\Xc}_{n,j} &=& (c_3-n-j)\,(n-j)\,\tilde{\Xc}_{n-1,j} +(c_2-j)\,j\,\tilde{\Xc}_{n-1,j-1} \;,\quad \tilde{\Xc}_{0,0}=1\;.
\ea
We can rearrange this equation into groups that involve $n$\/- or $j$\/-changes only
\ba
0 &=& n\big[ (c_3-n)\,\tilde{\Xc}_{n-1,j} -\tilde{\Xc}_{n,j} \big] + j\big[ (c_2-j)\,\tilde{\Xc}_{n-1,j-1} -(c_3-j)\,\tilde{\Xc}_{n-1,j} \big] \;,
\ea
which suggests to solve by a factorization Ansatz $\tilde{\Xc}_{n,j}=f_n\,g_j$, leading to two one-parameter recurrences that correspond to nullifying the two square brackets. These can be solved easily as
\ba
f_n&=&(c_3-n)\,f_{n-1}=\Big(\prod_{k=1}^n(c_3-k)\Big)\,f_0=\frac{\Gamma(c_3)}{\Gamma(c_3-n)}\,f_0 \;,\\
g_j&=&\frac{(c_2-j)}{(c_3-j)}\,g_{j-1}=\Big(\prod_{k=1}^j\frac{c_2-k}{c_3-k}\Big)\,g_0=\frac{\Gamma(c_2)}{\Gamma(c_2-j)}\,\frac{\Gamma(c_3-j)}{\Gamma(c_3)}\,g_0 \;.
\ea

Equation~\nr{eq:guts} is therefore solved by 
\ba\la{eq:gutsSoln}
\bar{\Xc}_{n,j} &=& \bin{n}{j}\,\frac{\Gamma(c_3)}{\Gamma(c_3-n)}\,\frac{\Gamma(c_3-j)}{\Gamma(c_3)}\,\frac{\Gamma(c_2)}{\Gamma(c_2-j)}\,\bar{\Xc}_{0,0} \;.
\ea
Reversing the definitions, replacing the $c_i$, and simplifying, we finally obtain the coefficients of \eq\nr{eq:ansatz} in closed form as
\ba\la{eq:ansatzSoln}
\Xc_{n,j}(d,\nu) &=& \frac{(-1)^j\,\po{\nu}{j}\,\po{1+n-\frac{d}2}{\nu}}{n!\,j!\,(n-j)!\,\po{1-\frac{d}2}{j+\nu}} \;.
\ea

With \eqs\nr{eq:ansatzSoln} and \nr{eq:ansatz} established as solution to \eq\nr{eq:T94b}, 
we can as well simply postulate this solution right away as in \eq\nr{eq:Baac}, and prove it via induction over $N$, 
see the comments in the main text. 
We think, however, that it is instructive to see the derivation as exposed here, given that a non-trivial two-dimensional 
recurrence with variable coefficients has been solved explicitly. 

In summary, at the core of the IBP relation \eq\nr{eq:T94b} lies the recurrence \eq\nr{eq:guts}, with closed-form solution \eq\nr{eq:gutsSoln}.

%
\section{Special mass case $[m,m,2m]$}
\la{se:112}

Employing the basic single-step IBP relation \eq\nr{eq:T92} for the special-mass case  $m_1=m_2\equiv m$, the recursion ends at
\ba \la{eq:B1a}
\hB^\nabc(d) &=&
\frac{\hr_1^\nabc(d)\,\hB^{1,1,0}(d)\;+\;\hr_2^\nabc(d)\,\hB^{1,0,1}(d)}{[m^2]^{\nu_1+\nu_2+\nu_3-2}}
\;,
\ea
where the $\hr_i$ are rational functions in the variable $d$ (there are only two master integrals here, due to the symmetry $\hB^{0,\nu_2,\nu_3}(d) = \hB^{\nu_2,0,\nu_3}(d)$), symmetric in their first two indices. 
For example,
\ba\la{eq:234}
\hr_1^{2,3,4}(d) &=& -\frac{(d-8)(d-6)(d-4)(d-2)}{(d-15)(d-13)(d-11)(d-9)}\, \frac{(d-5)(d^2-9d+6)}{258} \;,\\
\hr_2^{2,3,4}(d) &=& -\frac{(d-8)(d-6)(d-4)(d-2)}{(d-15)(d-13)(d-11)(d-9)}\, \frac{(d^5-56d^4+1223d^3-12916d^2+65220d-124560)}{393216}
\nonumber
\ea
are readily generated by an implementation of the recursion \nr{eq:T92}, specialized to this case.

While we do not presently know how to directly solve the 3-dimensional recurrence \eq\nr{eq:T92}, even in this simpler case in which all mass dependence can be pulled out of the coefficients, an analytic clue comes from \eq(3.4) of \cite{Davydychev:1992mt}, where the integral $\intB^\nabc_{m,m,M}(d)$ had been expressed as a sum of two terms, each being a product of Gamma functions and a hypergeometric function $\Fdc$ at argument $z=\frac{M^2}{4m^2}$. In the case $M=2m$ of interest to us here, we have $z=1$ and read off 
\ba \la{eq:4F3}
\hr_1^\nabc(d) &=& 
\frac{1}{\Gamma^2(1-d/2)}\, 
\frac{\Gamma(-e)\Gamma(\nu_1+e)\Gamma(\nu_2+e)\Gamma(a+e)}{\Gamma(d/2)\Gamma(\nu_1)\Gamma(\nu_2)\Gamma(\nu_1+\nu_2+2e)}\,
\hypA \;,\nonumber\\
\hr_2^\nabc(d) &=& 
\frac{4^{1-\nu_3}}{\Gamma^2(1-d/2)}\, 
\frac{\Gamma(a)\,\Gamma(e)}{\Gamma(\nu_1+\nu_2)\,\Gamma(\nu_3)}\,
\hypB 
\;,
\ea
with shorthands $a\equiv\nu_1+\nu_2-d/2$ and $e\equiv\nu_3-d/2$.
While this could count as a closed-form result, it is not terribly practical yet, since the $\Fdc$ are infinite sums.

To proceed, we can exploit the fact that we deal with strictly positive integers $\nu_i$ only, which allows us to employ the reduction
\ba\la{eq:PFQ}
{}_{p+1}F_{q+1}\hargs{a_1,\,\dots,\;a_p,\;c+k}{b_1,\,\dots,\;b_q,\;c}{z}
&=&
\sum_{\ell=0}^k \bin{k}{\ell}\,  
\frac{(a_1)_\ell\, \cdots (a_{p})_\ell\; z^\ell}
{(b_1)_\ell\, \cdots (b_{q})_\ell\; (c)_\ell}\;
{}_{p}F_{q}\hargs{a_1+\ell,\,\dots,\;a_p+\ell}{b_1+\ell,\,\dots,\;b_q+\ell}{z}
\ea
for some positive integer $k$, in order to reduce the order of the hypergeometric function.
For each of the $\Fdc$ of \eq\nr{eq:4F3} we can in fact identify {\em two} upper and lower parameter pairs that differ by such an integer $k$, leading to a reduction chain $\Fdc\rightarrow\Fcb\rightarrow\Fba$. The latter hypergeometric sum at unit argument then reduces to Gamma functions according to 
\ba\la{eq:E11}
\Fba\hargs{a,\;b}{c}{1} &=& \frac{\Gamma(c)\, \Gamma(c-a-b)}{\Gamma(c-a)\, \Gamma(c-b)} \;.
\ea

In order to make explicit how this reduction works in our case, we introduce the shorthand notation 
\ba
x_{12}\equiv{\rm max}(\nu_1,\nu_2) 
\quad,\quad 
n_{12}\equiv{\rm min}(\nu_1,\nu_2)
\quad(\;\Rightarrow\;\; x_{12}+n_{12}=\nu_1+\nu_2) \;,\\
c_{12} \equiv \lceil (\nu_1+\nu_2)/2 \rceil 
\quad,\quad 
f_{12} \equiv \lfloor (\nu_1+\nu_2)/2 \rfloor +\tfrac12
\quad(\;\Rightarrow\;\; c_{12}+f_{12}=\nu_1+\nu_2+\tfrac12) \;,
\ea
which we use to rewrite some of the parameters of the $\Fdc$ in \eq\nr{eq:4F3} in a more useful way. Exploiting the fact that the pair $\{\nu_1,\nu_2\}$ can be replaced by $\{n_{12},x_{12}\}$ (where both pairs are understood to be unordered sets, such as the parameter sets of the hypergeometric functions)
as well as $\{\frac{\nu_1+\nu_2}2,\frac{\nu_1+\nu_2+1}2\}\rightarrow\{c_{12},f_{12}\}$, we rewrite the two $\Fdc$ of \eq\nr{eq:4F3} as
\ba
\hypA &=& \hypAA \;,\\
\hypB &=& \hypBB \;.
\ea
In both cases the reduction formula \eq\nr{eq:PFQ} can now be applied twice, pairing the respective last values of the upper and lower parameter sets (the positive integers $k$ being $n_{12}-1$, $x_{12}-c_{12}$ and $\nu_3-1$). This immediately leads to a double sum over Gauss hypergeometric functions $\Fba$ at unit argument, which in turn are resolved by \eq\nr{eq:E11}. The coefficient functions of \eqs\nr{eq:4F3} are therefore 
\ba
\hr_1^\nabc(d) &=& \la{eq:E14}
\frac{1}{\Gamma^2(1-d/2)}\, 
\frac{\Gamma(-e)\Gamma(\nu_1+e)\Gamma(\nu_2+e)\Gamma(a+e)}{\Gamma(d/2)\Gamma(\nu_1)\Gamma(\nu_2)\Gamma(\nu_1+\nu_2+2e)}\,
\sum_{\ell=0}^{n_{12}-1} \bin{n_{12}-1}{\ell}
\sum_{\ell'=\ell}^{x_{12}-c_{12}+\ell} \bin{x_{12}-c_{12}}{\ell'-\ell}
\times\nonumber\\&&\times\;
\frac{\po{x_{12}+e}{\ell}}{\po{1+e}{\ell}}\,
\frac{\po{\nu_3}{\ell'}\,\po{a+e}{\ell'}}{\po{c_{12}+e}{\ell'}}\,
\frac{\Gamma(f_{12}+e)\,\Gamma(f_{12}-a-\nu_3-\ell')}{\Gamma(f_{12}+e-\nu_3)\,\Gamma(f_{12}-a)}
\;,\\
\hr_2^\nabc(d) &=& \la{eq:E15}
\frac{4^{1-\nu_3}}{\Gamma^2(1-d/2)}\, 
\frac{\Gamma(a)\,\Gamma(e)}{\Gamma(\nu_1+\nu_2)\,\Gamma(\nu_3)}\,
\sum_{\ell=0}^{\nu_3-1} \bin{\nu_3-1}{\ell}
\sum_{\ell'=\ell}^{x_{12}-c_{12}+\ell} \bin{x_{12}-c_{12}}{\ell'-\ell}
\times\nonumber\\&&\times\;
\frac{\po{x_{12}}{\ell}}{\po{1-e}{\ell}}\,
\frac{\po{n_{12}}{\ell'}\,\po{a}{\ell'}}{\po{c_{12}}{\ell'}}\,
\frac{\Gamma(f_{12})\,\Gamma(f_{12}-a-n_{12}-\ell')}{\Gamma(f_{12}-n_{12})\,\Gamma(f_{12}-a)}
\;.
\ea
As a quick check on these expressions, the example given in \eq\nr{eq:234} is readily reproduced.
On the other hand, we can use  \eq\nr{eq:newconj2} to read off a representation in terms of the $c^{(\Nu)}(d)$ for the coefficient functions 
$\hr_i$ of \eq\nr{eq:B1a} above ($\Nu=\nu_1+\nu_2+\nu_3$ as before):
\ba
\la{eq:newr32}
\hr_1^\nabc(d) &=& 
(-1)^\Nu \sum_{j=1-\nu_1}^{\nu_2-1} c^{(\Nu)}_{\nu_1,\nu_2;j}(d) \;,\\
\hr_2^\nabc(d) &=&
\frac4{2^\Nu}\Big[
\sum_{j=1-\nu_1}^{\nu_3-1} (-\tfrac12)^j\,c^{(\Nu)}_{\nu_1,\nu_3;j}(d)
+\sum_{j=1-\nu_2}^{\nu_3-1} (-\tfrac12)^j\,c^{(\Nu)}_{\nu_2,\nu_3;j}(d) \Big] \;.\qquad
\ea
Taking our solution for the $c^{(\Nu)}(d)$ from \eq\nr{eq:kallen3} and comparing this with \eqs\nr{eq:E14}, \nr{eq:E15}, 
we find full agreement (we have performed the comparison for a large set of integer values for the indices $\nu_1,\dots\nu_3$), 
providing a further independent check on our general solution \eq\nr{eq:kallen3}.

%
\section{Alternative derivation of \eq\nr{eq:gResult}}
\la{se:gAlt}

Here we would like to offer an alternative derivation of the main result of \se\ref{se:4.2}.

As we have seen, the analytically known special-mass case $\tB^\nabc(d)$ of \se\ref{se:B0mm}, when paired with the 
conjecture \eq\nr{eq:newconj2}, implies the constraint \eq\nr{eq:new0mm} on the sum of rational coefficient functions 
$c^{(\Nu)}(d)$. Using now \eq\nr{eq:kallen} to rewrite the $c^{(\Nu)}(d)$ in terms of the integers $G$, this constraint reads
\ba\la{eq:constraint}
\beta^\nabc(d) &=& 
\sum_{j=1-\nu_2}^{\nu_3-1} \sum_{k={\rm max}(1+j,1)}^{{\rm min}(\nu_2+j,\nu_3)} 
\frac{\po{1-\frac{d}2}{n_j-1}\,\po{1-\frac{d}2}{n_j-j-1}}{\po{1\!-\!\frac{d}2}{n_j-k}\,\po{\frac{d+3}2\!-\!\Nu}{n_j-k}}\,
\frac{(-1)^{\Nu+j}\,G_{\nu_2+j-k,\nu_3-k,\nu_1-1}}{(-4)^{n_j-k}\,\Gamma(k)\,\Gamma(k-j)} \;,\qquad
\ea
with $\Nu=\nu_1+\nu_2+\nu_3$, $n_j=\ceil{\frac{\Nu+j}2}$, and $\beta$ given in terms of Gamma functions in \eq\nr{eq:B0mm}.

Both sides of \eq\nr{eq:constraint} are rational functions in $d$, so have to agree in particular in their pole structure.
Analyzing the rational function on the right-hand side, we see that there are only single poles in odd dimensions, i.e.\ $\frac1{d-p}$ with odd integers $p$, 
which arise from the second of the Pochhammer symbols in the denominator, and where $p=2\Nu-1-2\ell$ with $\ell\in\{1,\dots,n_j-k\}$.
Focusing on the unique pole with the {\em smallest} such $p$, we need to look for summation parameters that maximize the difference $(n_j-k)$.
Since $n_j$ is defined by a ceiling function, we need to distinguish two cases.
For {\em even} values of $\Nu$, the two terms $j=0$, $k=1$ and $j=1$, $k=2$ both maximize $(n_j-k)=\frac\Nu2-1$ 
which leads to a 'minimal' pole at $p=\Nu+1$, with all other terms of the double sum giving smaller values for $(n_j-k)$ and hence poles at larger $p$.
For {\em odd} values of $\Nu$, only one term contributes to this unique pole: $j=0$, $k=1$ has $(n_j-k)=\frac{\Nu+1}2-1$ 
and leads to a 'minimal' pole at $p=\Nu$, with all other terms of the double sum again leading only to poles at larger $p$.

Let us now look at \eq\nr{eq:constraint} in the case of odd $\Nu$ for simplicity (whence $n_j=\frac{\Nu+1+j}2$), 
in particular at the residues of the unique single pole $\frac1{d-\Nu}$. In practice, we set $d=\Nu+\ep$ and keep only the divergent term when $\ep\rightarrow0$. As analyzed above, the double sum collapses to a single term $j=0$, $k=1$, giving
\ba
\beta^\nabc(d=\Nu+\ep) &\stackrel{{\rm odd}\;\Nu}=& 
\frac{\po{1-\frac\Nu2}{\frac{\Nu-1}2}}{\po{\frac{3-\Nu}2}{\frac{\Nu-3}2}}\,
\frac{(-2)\,G_{\nu_2-1,\nu_3-1,\nu_1-1}}{(-4)^{\frac{\Nu-1}2}}\,
\Big[\frac1\ep+{\cal O}(\ep^0)\Big] \;.
\ea

According to \eq\nr{eq:G} we have $G_{a_2,a_3,a_1}=2g_{a_1,a_2,a_3}$, which allows us to constrain the unknown sequence of integers as 
(odd $\Nu$ implies even index sum $a_1+a_2+a_3$, such that we can use the integer $A\equiv\frac{a_1+a_2+a_3}2$ for brevity)
\ba
g_{a_1,a_2,a_3} &=&
\frac{\po{-A}{A}\,(-4)^A}{\po{-A-\frac12}{A+1}}\,\Big[\lim_{\ep\rightarrow0}\ep\,\beta^{a_1+1,a_2+1,a_3+1}(d=2A+3+\ep)\Big] \nonumber\\&=&
\frac{4^A\,A!}{a_1!\,a_2!\,a_3!\,\po{\frac12}{a_1-A}\,\po{\frac12}{a_2-A}\,\po{\frac12}{a_3-A}} 
\;.\la{eq:gResult2}
\ea
In the last step we have used \eq\nr{eq:B0mm} and performed the indicated limit (in which $\beta$ turns out to be fully symmetric in the $a_i$), used that $\po{-A}{A}=(-1)^A\,A!$ on the integers $A$, 
and employed the Euler reflection formula $\Gamma(\tfrac12+n)\,\Gamma(\tfrac12-n)=(-1)^n\,\Gamma^2(\tfrac12)$ for integers $n=A-a_i$ and $n=A+1$.

Equation \nr{eq:gResult2} coincides with \eq\nr{eq:gResult}, and provides an independent check of its validity.

%

\end{document}